\documentclass[aps,prb,twocolumn,groupedaddress]{revtex4}

\usepackage{graphicx}
\usepackage{dcolumn}
\usepackage{bm}
\usepackage{longtable}
\usepackage{subfigure}

\newcommand{\kjo}[1]{\rm{$^{#1}$}}
\newcommand{\kjn}[1]{\rm{$_{#1}$}}
\newcommand{\kj}[1]{$\rm{#1}$}

\begin{document}


\title{Bonding in MgSi and AlMgSi Compounds Relevant to AlMgSi Alloys}


\author{Anders G. Fr\o seth}
\email[Email: ]{Anders.Froseth@phys.ntnu.no}
\affiliation{Norwegian University of Science and Tecnology (NTNU), N-7034 Trondheim, Norway}
\author{Peter M. Derlet}
\affiliation{Paul Scherrer Institute, CH-5232 Villigen PSI, Switzerland}
\author{Sigmund J. Andersen}
\author{Calin D Marioara}
\affiliation{SINTEF Materials Technology, Applied Physics, N-7050 Trondheim, Norway}
\author{Ragnvald H\o ier}
\affiliation{Norwegian University of Science and Tecnology (NTNU), N-7034 Trondheim, Norway}


\date{\today}

\begin{abstract}
The bonding and stability of MgSi and AlMgSi compounds relevant to
AlMgSi alloys is investigated with the use of (L)APW+(lo) DFT
calculations. We show that the $\beta$ and $\beta''$ phases found
in the precipitation sequence are characterised by the presence of
covalent bonds between Si-Si nearest neighbour pairs and
covalent/ionic bonds between Mg-Si nearest neighbour pairs. We
then investigate the stability of two recently discovered
precipitate phases, U1 and U2, both containing Al in addition to
Mg and Si. We show that both phases are characterised by tightly
bound Al-Si networks, made possible by a transfer of charge from
the Mg atoms.
\end{abstract}


\maketitle


\section{Introduction}
Precipitation or age hardened alloys are today one of the most
important alloy types in industry. In the AlMgSi alloy system,
Mg-Si and Al-Mg-Si precipitates formed during specific heat
treatments give rise to a very significant increase in strength.
The precipitation sequence is generally accepted to be:
\begin{equation}
SSSS \rightarrow \text{Mg/Si Clusters} \rightarrow GPZ \rightarrow
\beta'' \rightarrow \beta' \rightarrow \beta,
\end{equation}
where GPZ refers to a Guinier Preston Zone and SSSS refers to a
Super Saturated Solid Solution. Very little is known about the
early stages of the precipitation process. However, it is believed
that when the SSSS is heat treated Mg and Si atoms quickly diffuse
substitutionally to form small clusers due to the large amount of
quenched-in vacancies \cite{porter} (a large part of the vacancies
move to interfaces like surfaces and grain boundaries in the later
stages of the heat treatment). Although the details are hard to
investigate experimentally, several studies of such clustering
have been carried out using atom probe microscopy
\cite{atomprobe1, atomprobe2}. The first phase which can be
resolved using high resolution electron microscopy (HREM) is the
GPZ. From this, a model of its crystal structure has recently been
proposed \cite{calin}. The structure of the proceeding phase,
$\beta''$, was also solved using electron microscopy
techniques\cite{science}, a result which has later been supported
by \emph{ab initio} calculations\cite{peter}.

Earlier it was believed that the GPZ, $\beta''$ and $\beta$ phases
all had the stoichiometry of the $\beta$ phase, \kj{Mg_{2}Si}, and
that alloys should be optimized accordingly. Thus, the terminal
$\beta$ phase was of primary importance. Later however, it had
been confirmed that the $\beta''$ gives a greater contribution to
the hardness due to its semi-coherent interface with the aluminium
matrix and needle-like shape, which is more effective for
dislocation pinning \cite{betadphard}. The $\beta''$ stoichiometry
has been shown to be \kj{Mg_{5}Si_{6}}\cite{science}.

Recently, several additional phases have been identified
experimentally\cite{matsuda, U1experimental, U2experimental},
giving the extended precipitation sequence:
\begin{eqnarray}
SSSS \rightarrow Mg/Si \text{ Clusters} \rightarrow GPZ \rightarrow \beta'' \\
\rightarrow (\beta' + U2 + U1 + B') \rightarrow \beta. \nonumber
\end{eqnarray}
In ref. \onlinecite{matsuda} the phases $U1$, $U2$ and $B'$ are
referred to as type \emph{A}, \emph{B} and \emph{C} respectively.
These three phases, in addition to $\beta'$, are often grouped
together since little is known about their interdependence.
However it is believed that the peak of concentration with respect
to time for each of these structures follows the ordering given by
the above precipitation sequence \cite{matsuda}. They all form
relatively late in the precipitation sequence usually at
temperatures in the range 200-300 \kj{^{\circ}C}, and in Si-rich
alloy compositions.

It has been considered a general rule of thumb that successful
aluminium precipitation hardening alloys contain secondary and
ternary alloying elements which are larger and smaller than
aluminium \cite{sizerule}. Now, the concept of atom size in this
context must be based on the type of bonding involved. One can
have either ionic, metallic or covalent radii for the constituent
elements giving dramatically different values for atomic size
\cite{kittel}. It is clear that when studying the electronic
density of compound structures this type of concept may lead to an
oversimplification. It it therefore interesting to carry out a
theoretical study of the electronic structure and bonding
characteristics of the relevant phases with respect to their
relative stability. This is the purpose of the present work, where
we employ full potential \emph{ab initio} methods based on Density
Functional Theory (DFT) to investigate the bonding within the
above mentioned structures. In section II we describe the Linear
Augmented Plane Wave + local orbitals approach, (L)APW+(lo), used for
all calculations. Section III describes the results obtained for
the three models of precipitate phases containing only Mg and Si
($\beta''$, $\beta'$, and $\beta$), and section IV deals with the
phases containing Al, Mg and Si ($U1$ and $U2$).


\section{Method}
The \emph{ab initio} calculations were performed using WIEN2k, a
program package implementing the full potential (L)APW+(lo) DFT
method\cite{wien2k}. The Augmented Plane Wave approach (APW)+(lo)
method\cite{singh} differs from the LAPW method in the
linearization of the APW's. In slater's original APW method
\cite{slater} the unit cell is partitioned into non-overlapping
atomic spheres and an interstitial region.
The basis functions are plane waves for the interstitial region
and radial wave functions within the atomic spheres. In the LAPW
method the basis functions inside the spheres are linearized with
respect to the energy $E_l$:
\begin{equation}
\psi_{\bm{k_n}}(\bm{r}) = \sum_{lm}(A_{lm,\bm{k_n}}u_l(r,E_l) + B_{lm,\bm{k_n}}\dot{u}_l(r,E_l))Y_{lm}(\bm{\hat{r}})
\end{equation}
where $u_l(r,E_l)$ are radial functions and $Y_{lm}(\bm{r})$ are spherical harmonics. $\dot{u}_l(r,E_l)$ is
the energy derivative of $u_l(r,E_l)$. %
$A_{lm,\bm{k_n}}$ and $B_{lm,\bm{k_n}}$ are determined by matching
the above basis to the value and derivative of the plane waves for each k-vector at the sphere boundary.
An alternative way for doing the linearization is the APW+lo method. One starts with the original APW's and
adds local orbitals to obtain the variational flexibility in the basis functions.
\begin{eqnarray}
\psi_{\bm{k_n}}(\bm{r}) = \sum_{lm}(A_{lm,\bm{k_n}}u_l(r,E_l) + \phi_{lo})Y_{lm}(\bm{\hat{r}}) \\
\phi_{lo} =  B_{lm}u_l(r,E_l) + C_{lm}\dot{u}_l(r,E_l)
\end{eqnarray}
At first sight this looks very similar to the LAPW basis. However,
the $B_{lm}$ and $C_{lm}$ are no longer dependent on the
wave-vector and are determined by the requirement that the local
orbital is zero at the sphere boundary and normalized. The great
advantage of this scheme is that the calculations converge to
results almost identical with those of the LAPW method but for
dimensioning parameters which effectively leads to a smaller basis
set \cite{madsen}. The WIEN2k code uses a mixed APW+lo/LAPW basis
set, exploiting the advantages of both methods.

To make the results for the different structures comparable, we
used the same set of APW+(lo) parameters for all calculations:
$R_{mt}=2.1$ Bohr, $R_{mt}K_{max} = 7$ and $G_{max} = 14$
\kj{Ry^{1/2}}. Here, $R_{mt}$ is the muffin tin radius, $K_{max}$
is the plane wave cut-off and $G_{max}$ is the maximum Fourier
component of the electron density. For all calculations we used
the modified tetrahedron method\cite{tetra} for Brillouin zone
integrations. All k-point meshes were checked for convergence.
Thus in general, the highly symmetric structures (like bulk Al, Si
and Mg) with few symmetrically inequivalent atoms in the unit
cell, require a denser k-point mesh than the precipitate phases
with larger unit cells and a lower symmetry. For the
exchange-correlation potential we used the Generalized Gradient
Approximation (GGA) of Perdew et. al \cite{GGA}

\section{\kj{MgSi} Phases}

\subsection{$\beta$}
\begin{figure}
\centerline{\rotatebox{00.00}{\scalebox{0.50}{\includegraphics{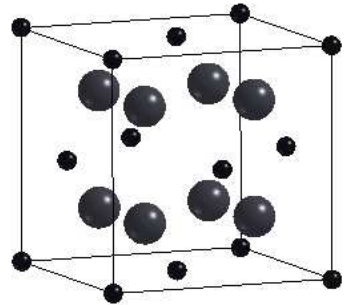}}}}
\caption{\label{Fluorite} Conventional unit cell for fluorite \kj{Mg_2Si}, equivalent to
the $\beta$ phase in the AlMgSi precipitaion sequence. The large spheres are Mg
atoms and the smaller ones are Si.}
\end{figure}

The $\beta$ phase (fluorite \kj{Mg_2Si}) is the terminal
equilibrium structure of the precipitation sequence. It has a fcc
primitive unit cell (space group \kj{Fm\bar{3}m} (225)), with an
experimental lattice parameter a=6.39 {\AA} \cite{calin}. It forms
precipitates of a plate-like or cubic shape up to 20 \kj{\mu m} in
diameter. Its interface with the Al matrix is fully incoherent
\cite{calin}. The conventional unit cell, containing 8 Mg atoms
and 4 Si atoms,  is shown in fig.~\ref{Fluorite}. As can be seen,
each of the Si atoms in the structure has 8 Mg nearest neighbours,
giving each Mg atom 4 Si nearest neighbours at the same distance.
The Si atoms are arranged as an fcc lattice interpenetrated by a
sc Mg lattice.

Several \emph{ab initio} studies of the bonding in Fluorite
\kj{Mg_2Si} have been carried out in the past \cite{frenchbeta1,
frenchbeta2, betawood}.  However, to our knowledge, none of these
involved the LAPW or (L)APW+(lo) DFT method.  For our calculations
we use the experimentally observed lattice constant for this phase
in the Al matrix: 6.39 \AA. Performing a volume relaxation, this
value differed by only 0.25 \% from the calculated optimized value
6.37 {\AA}, and by 0.6 \% from the experimental value for bulk
\kj{Mg_2Si}, 6.338 \AA \cite{betaexp}. The calculated bulk
modulus, derived from a second order Birch fit, was 54.3 GPa,
compared to 59 GPa from experiment \cite{betaexp}. Using the 6.39
{\AA} lattice constant the nearest neighbour (nn) Mg-Si distance
is 2.77 {\AA}, the Si-Si nn distance is 4.51 {\AA} and the Mg-Mg
nn distance is 3.29 {\AA}.

\begin{figure}
\centerline{\rotatebox{00.00}{\scalebox{0.85}{\includegraphics{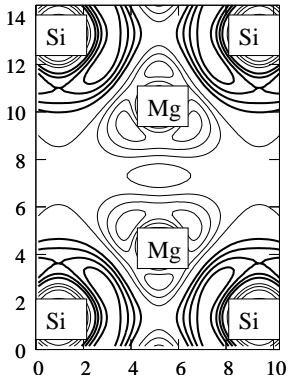}}}}
\caption{\label{Mg2Si3D} Bonding charge density for $(\bar 101)$
plane. The two central peaks are Mg atoms. The four corner peaks are
Si atoms. Thick lines represent positive contour lines and thin lines
represent negative contour lines.}
\end{figure}

Past linear-combination-of-atomic-orbitals Hartree Fock and DFT
pseudopotential calculations \cite{frenchbeta1, betawood} indicate
considerable charge transfer from the electropostive Mg atoms to
the electronegative Si atoms resulting in a partly ionic Si-Mg
bond for the $\beta$ phase. This result is also supported by the
present work. Fig.~\ref{Mg2Si3D} displays the bonding charge
density for the $(\bar 101)$ plane of fig.~\ref{Fluorite}. We define
the bonding charge density as the difference between the converged
valence charge density from DFT and the charge density derived
from the isolated neutral atoms. Thus the bonding charge density
indicates the charge transfer resulting from the converged
electronic structure. In fig.~\ref{Mg2Si3D}, the thick lines
represent positive contours corresponding to charge transfer to
the region, while the thinner lines represent negative contours
corresponding to charge transfer away from the region. The
concentration of thick lines around the Si atoms thus displays a
net build up of charge around the Si atoms counterbalanced by a
net reduction of charge around the central Mg atoms, indicating
that some ionicity is indeed at play.

\begin{figure}
\centerline{\rotatebox{-90.00}{\scalebox{0.35}{\includegraphics{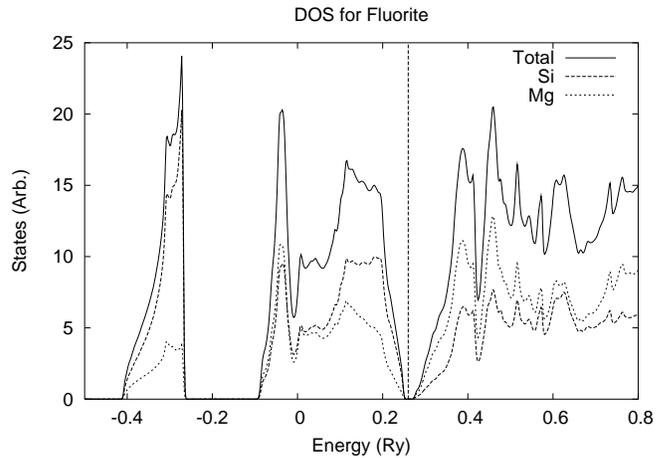}}}}
\caption{\label{Mg2SiDOS} Muffin tin decomposed DOS for the $\beta$
phase. Note that the atomic Mg DOS is half the magnitude of the total
Mg DOS. The vertical line represents the fermi level.}
\end{figure}

Fig.~\ref{Mg2SiDOS} displays the corresponding total and partial
electronic density of states (DOS) for the $\beta$ phase. Here the
partial DOS represents the ``Muffin tin decomposed'' DOS for both
the Mg and Si atoms, in which the occupied states are projected
onto the muffin tin eigenvectors (see section II) of the
particular atom. In this way, some information can be gained on
the local DOS at each atom, however such a procedure is not
complete since the interstitial region cannot be locally resolved.
On the other hand, the total DOS represents the DOS derived from
all the``Muffin tin decomposed'' DOS and the interstitial DOS of
the entire computational cell. Note that the calculation is based
on the primitive unit cell, containing 1 symmetrically
inequivalent Si atom and 2 symmetrically  inequivalent Mg atoms,
thus the partial Mg DOS contains the contribution for 2 Mg atoms,
and the Si DOS the contribution from 1 Si atom.

A general feature of the total and partial DOS in
fig.~\ref{Mg2SiDOS} is the two broad bands below the Fermi energy
(indicated by the horizontal line at 0.26 Ry) separated by an
energy of approximately 0.4 Rydbergs. That the occupancy is
dominated by Si states, where the partial Mg DOS represents the
two inequivalent Mg atoms, is a further indication of the presence
of ionicity whereby the Mg has simply donated electrons to the Si.
In addition this phase has a band-gap. The magnitude of the gap at
the $\Gamma$-point is 0.13 Ry (1.77 eV), which is somewhat smaller
than the experimental value of 2.27 eV for a lattice constant of
6.338 {\AA} \cite{tysktabell}.

\begin{figure}
\begin{tabular}{c}
\mbox{a)}{\rotatebox{-90.00}{\scalebox{0.30}{\includegraphics{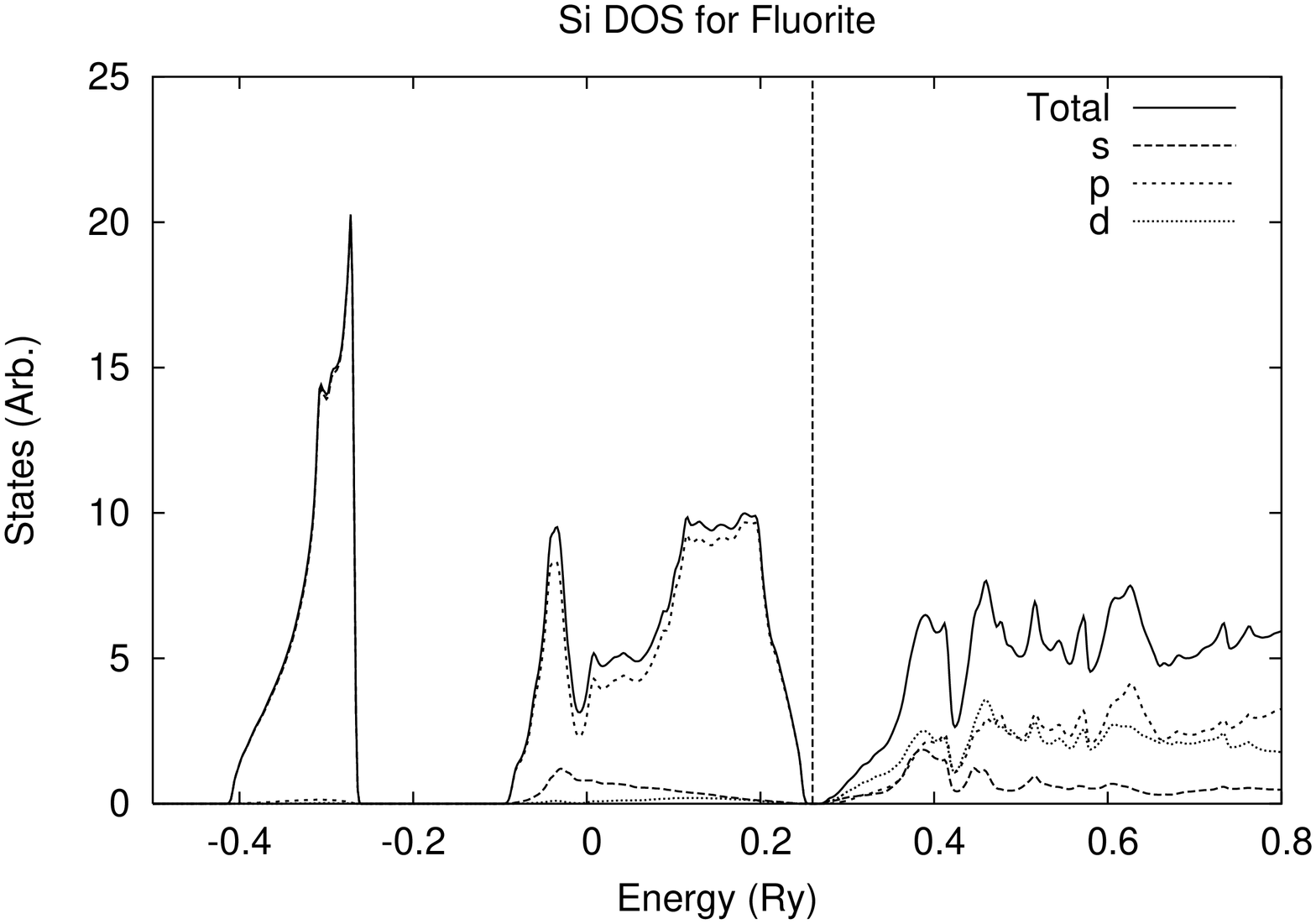}}}
} \\
\mbox{b)}{\rotatebox{-90.00}{\scalebox{0.30}{\includegraphics{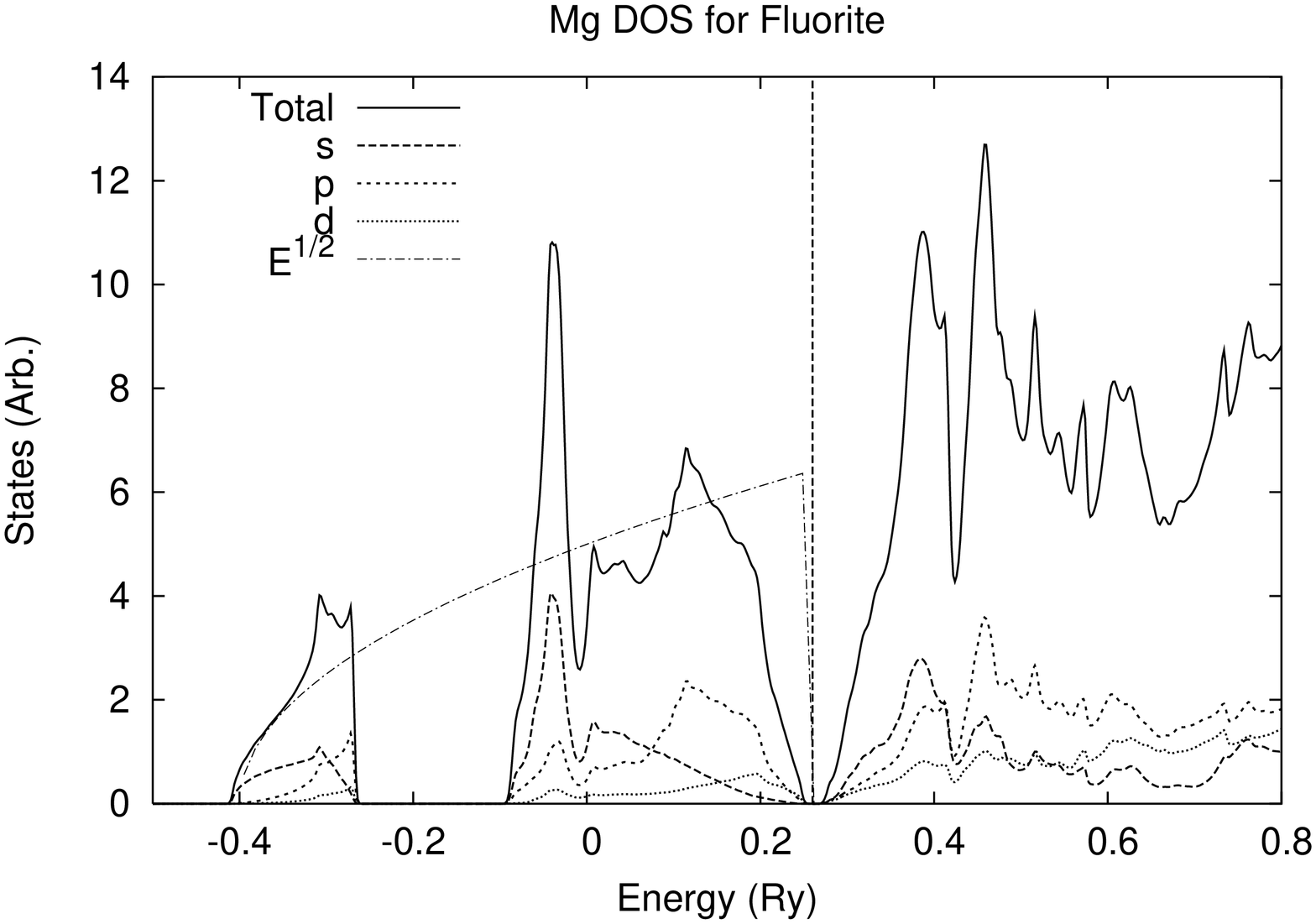}}}
}
\end{tabular}
\caption{\label{Mg2Si_DOS2}a) Si and b) Mg muffin tin decomposed DOS
for the $\beta$ phase. Note that for the Si DOS, the $s$-curve is
practically superimposed on the $total$-curve for the lower band in the
range -0.4 to -0.25 Ry. Scaled square-root curve shown for comparison.}
\end{figure}

Figs.~\ref{Mg2Si_DOS2}a and b now display the $s$, $p$ and $d$
decomposed partial DOSs for both Si and the two inequivalent Mg
atoms. In this case, the total DOS represents the total muffin tin
decomposed DOS for the particular atom, that is, the sum of the
$s$, $p$ and $d$ partial DOSs. Inspection of
fig.~\ref{Mg2Si_DOS2}a reveals that the lower band consists of $s$
states where as the upper band consists of predominantly $p$
states with a little $s$ state character. For the case of the Mg,
fig.~\ref{Mg2Si_DOS2}b, the DOS below the Fermi level largely
mimics that of Si apart from both occupied bands being of mixed
$s$ and $p$ character.

Insight into the general features of the DOSs shown in
fig.~\ref{Mg2Si_DOS2} can be understood from the perspective of
beginning with pure fcc Si with a lattice constant of 6.39{\AA},
giving a rather large nn Si-Si separation of 4.52{\AA}. For such a
system there can be no strong hybridization between the atomic
valence $s$ and $p$ states, maintaining a gap of approximately 0.3
Ry in the corresponding DOS (not shown). The $s$ and $p$ band
centers differ by about 0.4 Ry, which is not so dissimilar from
the 0.49 Ry difference between the isolated atom 3s and 3p valence
states. With the addition of the interpenetrating cubic array of
Mg atoms, both bands broaden reducing the gap to about 0.2 Ryd.
Thus in fig.~\ref{Mg2Si_DOS2}a, the dominant s character of the
filled lower band and the dominant $p$ character of the 
upper filled band arises from the strong onsite Si orthogonality
requirement, whereas in fig.~\ref{Mg2Si_DOS2} the mixed $s$ and
$p$ character of the corresponding Mg bands arises from the Mg-Si
$s$ and $p$ matrix interaction elements. We note that the
corresponding heights of the Mg total DOS are significantly less
than the Si DOS, following approximately the square-root
dependence of a metallic DOS. Thus, through charge transfer to the
Si, the Mg plays the role of ``strengthening'' the Si backbone
lattice, providing an explanation of the origin of the $\beta$
phase band gap through the pulling down of the excited $s$, $p$
and $d$ states of atomic Si.

\subsection{$\beta$''}
\begin{figure}
\begin{tabular}{c}
\mbox{a)}{\rotatebox{00.00}{\scalebox{0.35}{\includegraphics{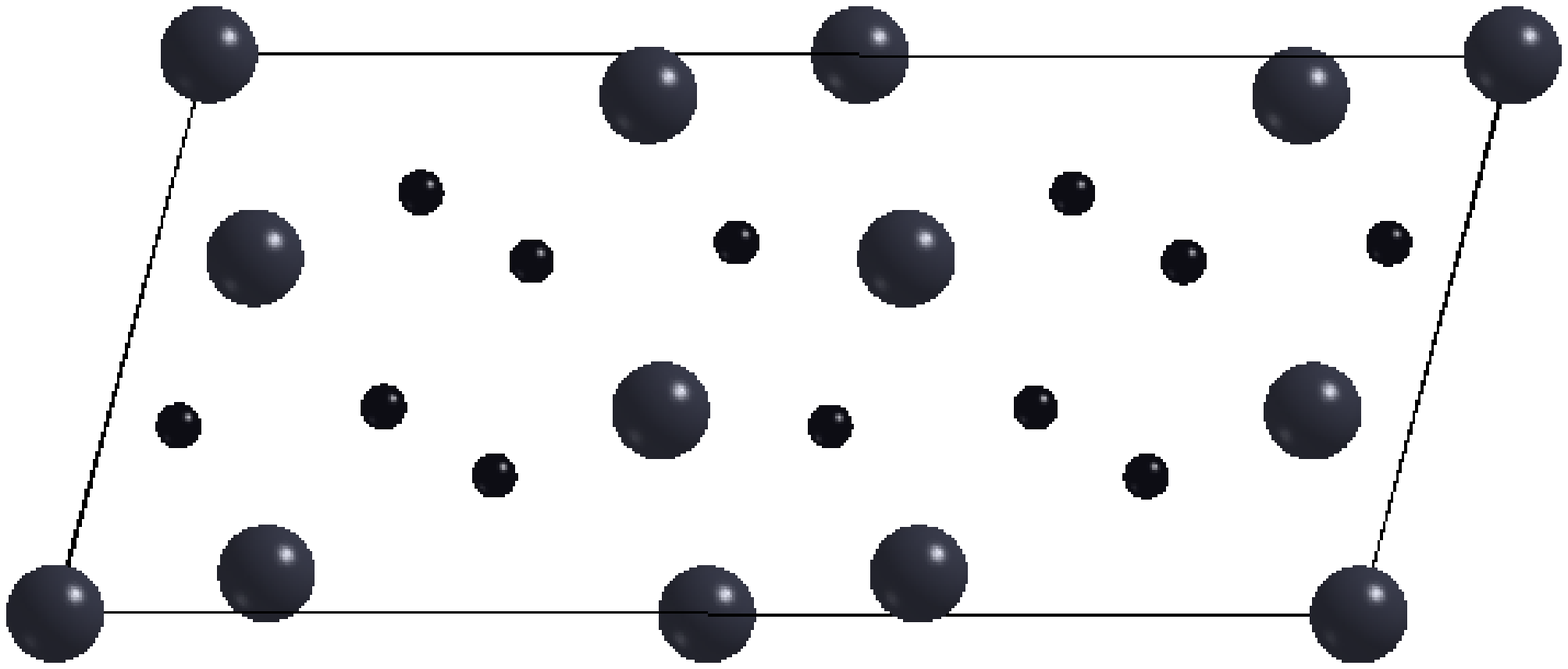}}} } \\
\mbox{b)}{\rotatebox{00.00}{\scalebox{0.35}{\includegraphics{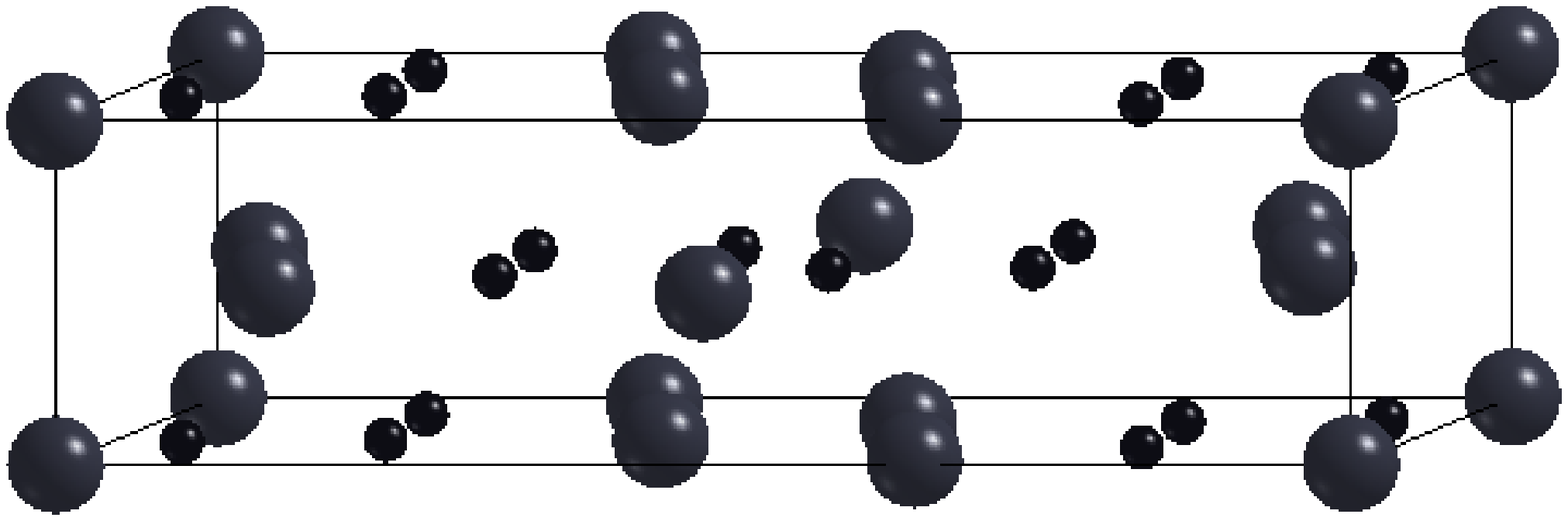}}} }
\end{tabular}
\caption{\label{betadpstruct} Conventional unit cell for $\beta''$ seen from a) \kj{<001>} direction, and  b)
slightly away from the \kj{<010>} cartesian direction. The large circles represent Mg and the small circles Si.}
\end{figure}

\begin{figure}
\begin{tabular}{l}
\rotatebox{0.00}{\scalebox{0.75}{\includegraphics{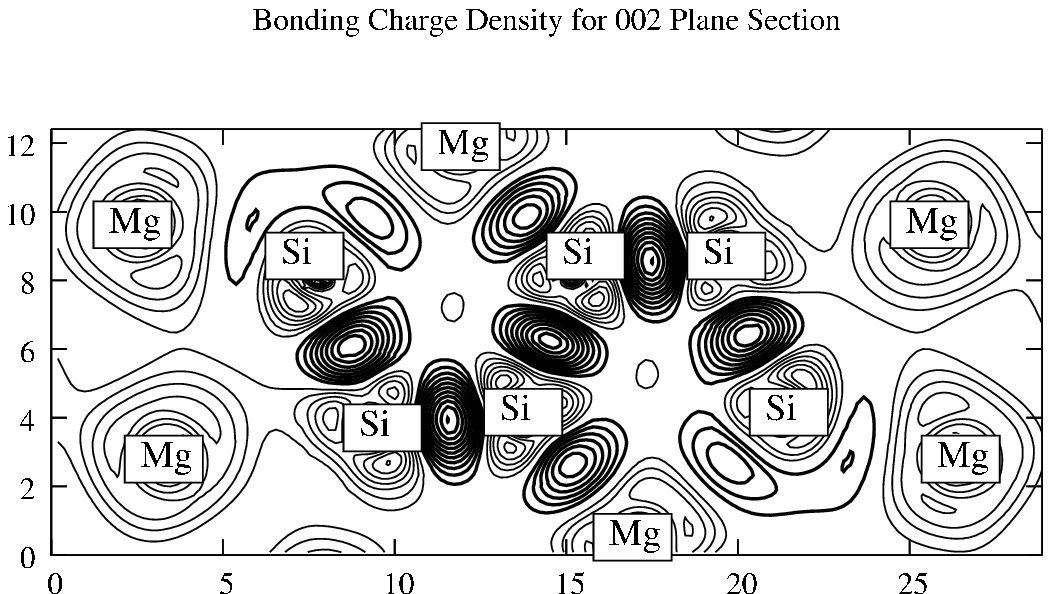}}}
\end{tabular}
\caption{\label{betadprho} Bonding charge density for (002) plane section of the $\beta''$ Structure.
Thick contour lines represent a positive charge transfer, while thin lines represent a negative charge transfer.}
\end{figure}

\begin{figure}
\centerline{\rotatebox{-90.00}{\scalebox{0.35}{\includegraphics{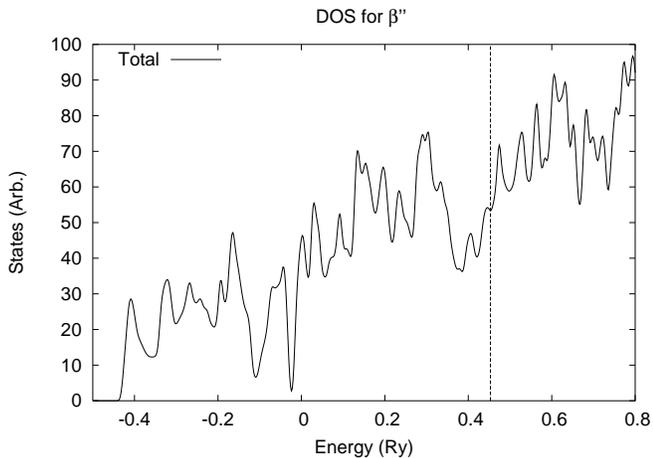}}}}
\caption{\label{betadptotalDOS} Total DOS for the $\beta''$ phase. The vertical line
represents the fermi level.}
\end{figure}

\begin{figure}
\begin{tabular}{c}
\mbox{a)}{\rotatebox{-90.00}{\scalebox{0.30}{\includegraphics{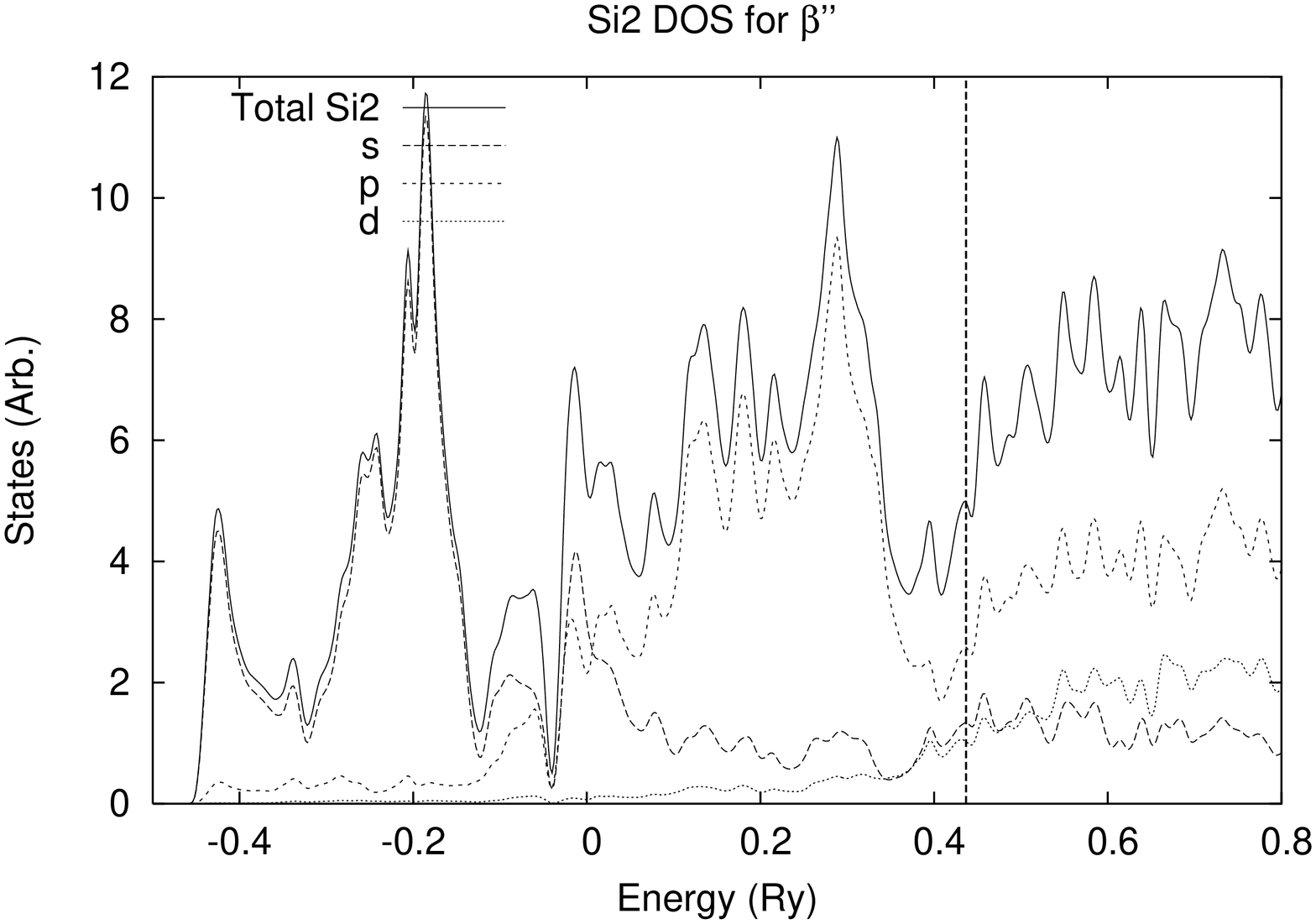}}}
} \\
\mbox{b)}{\rotatebox{-90.00}{\scalebox{0.30}{\includegraphics{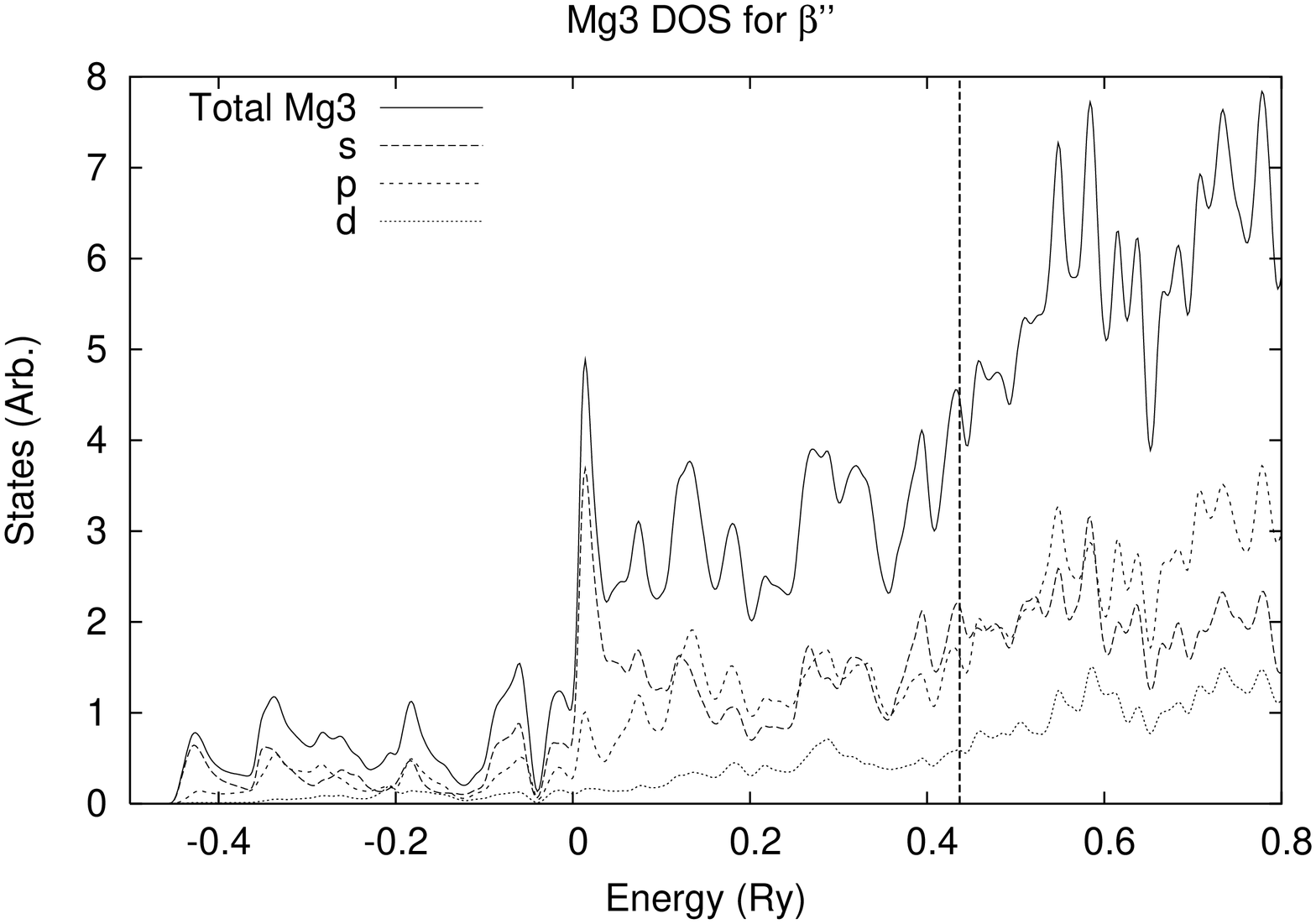}}}
}
\end{tabular}
\caption{\label{betadppsimgDOS} Partial DOS for the a) Si2 and b) Mg3 (see table. \ref{betadpcoordinates}) atom in the $\beta''$ phase.}
\end{figure}

The $\beta''$-phase has a base-centred-monoclinic conventional
unit cell (space group \kj{C2_m}) with experimental lattice
parameters $a=15.16$ {\AA}, $b=6.74$ {\AA}, $c=4.05$ {\AA}, and
$\gamma_{ab}$=\kj{105.3^{\circ}} \cite{science}.
Fig.~\ref{betadpstruct} displays the conventional cell for two
viewing orientations. This phase forms precipitates of a
needle-like shape, typically with a thickness of 30-50 {\AA} and a
length of 300-400 {\AA}. The needle length, which is along the
\kj{<001>} direction of the $\beta''$-phase unit cell, runs
parallel to the \kj{<001>} direction of the Aluminium matrix.  
For the present work we use the experimentally derived lattice 
parameters and the relaxed atomic positions of the inequivalent 
Si and Mg atoms by minimizing the forces by a Newton-Coates 
procedure with respect to the \kj{C2/m} space group. The resulting 
inequivalent atomic coordinates are listed in table 
\ref{betadpcoordinates}. The given space group of this phase was 
first solved by high resolution electron microscopy techniques 
\cite{science} and later supported by FLAPW DFT calculations \cite{peter}.
\begin{table}
\caption{\label{betadpcoordinates} Relaxed fractional coordinates for the inequivalent atomic positions
of the $\beta''$ phase.}
\begin{ruledtabular}
\begin{tabular}{lccr}
\bf{Atom} & \bf{a} & \bf{b} & \bf{c} \\
\hline
Mg1 & 0.0 & 0.0 & 0.0 \\
Mg2 & 0.346(1) & 0.071(9) & 0.0 \\
Mg3 & 0.421(6) & 0.063(6) & 0.0 \\
Si1 & 0.055(7) & 0.662(7) & 0.0 \\
Si2 & 0.194(4) & 0.250(5) & 0.0 \\
Si3 & 0.209(1) & 0.627(5) & 0.0 \\
\end{tabular}
\end{ruledtabular}
\end{table}
Fig.~\ref{betadprho} displays the bonding charge density contour
plot for the (002) plane. A dominant feature is the
concentration of charge between the Si nearest neighbours, and to
a lesser extent between the Si and Mg nearest neighbours,
indicating that covalency is at play in this system \cite{peter}.
Such charge transfer to the bonding regions originates from the
core regions of both the Si and Mg atoms, in addition to the
homogeneous interstitial region between the Mg atoms. The
depletion of charge from the Mg is comparable to that seen in the
$\beta$-phase, indicating that for this system both ionicity and
covalency is present in the bonding.  The nearest neighbour
distance range in this structure is 2.39-2.53 {\AA} for Si-Si and
2.61-2.84 {\AA} for Si-Mg. This is not dissimilar to the nearest
neighbour distance of 2.33 {\AA} in covalently bonded diamond
cubic Si and 2.76 {\AA} in the covalent/ionic equilibrium $\beta$
phase. Moreover, one of the inequivalent Si atoms is sp\kjo{3}
(tetrahedrally) coordinated and another is sp\kjo{2} coordinated
\cite{peter}.
Fig.~\ref{betadptotalDOS} displays the Total DOS (including the
muffin tin and interstitial regions) for the $\beta$ phase. Unlike
the $\beta$ phase, there is no band gap. However, depressions can
be identified at -0.1 and 0.4 Rydbergs reminiscent of the gaps
seen in fig.~\ref{Mg2SiDOS}. By inspection of the partial DOS for
each inequivalent atom we find that for the Si atoms there is a
dominant $s$ character for the lower energy states ($<-0.2$ Ry),
whereas, at about 0.2 Ry, the $p$ character dominates. In between
these regimes, a mixture of $p$ and $s$ character exists,
indicating strong hybridization. Indeed fig.~\ref{betadppsimgDOS}a
displays the partial DOS for Si2, which from past work is locally
sp\kjo{3} coordinated \cite{peter}, showing approximate
correspondence with the diamond cubic Si DOS \cite{liu_cohen}. In
particular the dominant $s$ character of the L\kjn{2'}, the $s$
and $p$ character of the L\kjn{1} and the dominant $p$ character
of X\kjn{4} the high symmetry points. Alternatively,
fig.~\ref{betadppsimgDOS}b shows the partial DOS for Mg3, which
from charge density profile analysis \cite{peter} is in a
homogeneous metallic environment, displaying an approximate
shifted metallic square-root behaviour.

\section{\kj{AlMgSi} Phases}

\subsection{$U1$}
\begin{table}
\caption{\label{U1coordinates} Fractional coordinates for the
inequivalent atomic positions of the $U1$ phase
\cite{U1experimental}.}
\begin{ruledtabular}
\begin{tabular}{lccr}
\bf{Atom} & \bf{a} & \bf{b} & \bf{c} \\
\hline
Mg & 0.0 & 0.0 & 0.0 \\
Al & 1/3 & 2/3 & 0.632(8) \\
Si & 1/3 & 2/3 & 0.243(8) \\
\end{tabular}
\end{ruledtabular}
\end{table}
\begin{figure}
\begin{tabular}{c}
\mbox{a)}{\rotatebox{00.00}{\scalebox{0.35}{\includegraphics{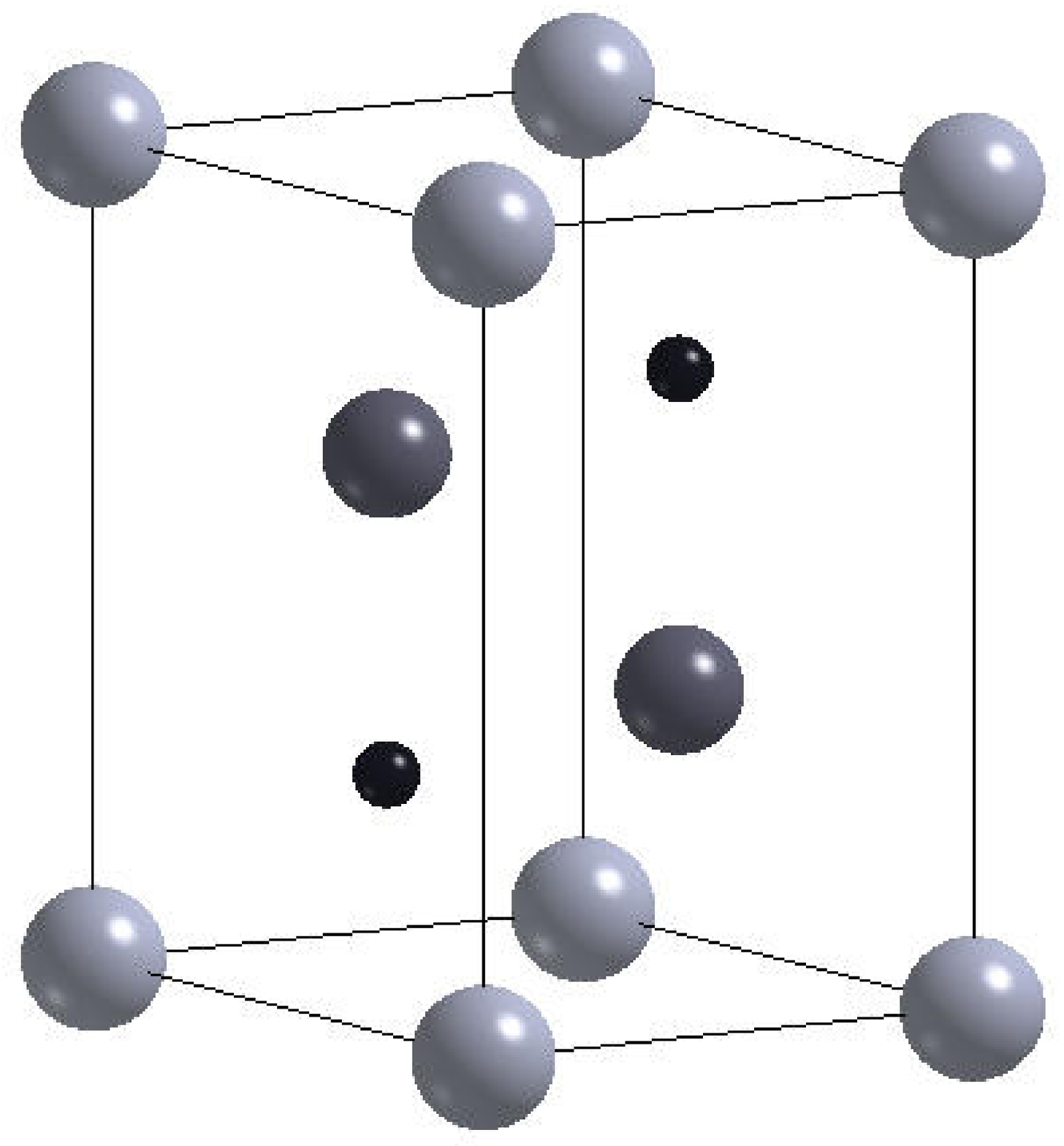}}} } \\
\mbox{b)}{\rotatebox{00.00}{\scalebox{0.35}{\includegraphics{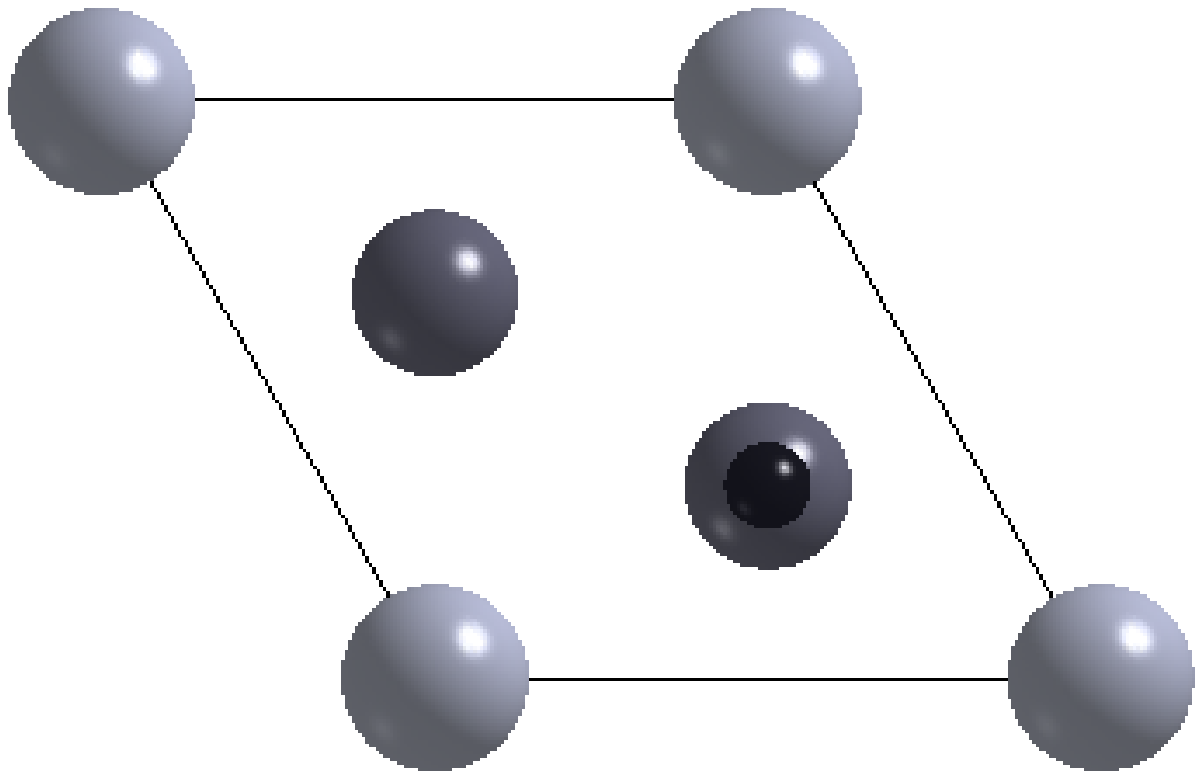}}} }
\end{tabular}
\caption{\label{U1new} $U1$ conventional unit cell in a)
perspective and b) \kj{<001>} direction. The small, black spheres are Si
atoms, the large dark spheres Al atoms, and the large white
spheres Mg atoms.}
\end{figure}
The $U1$ conventional/primitive unit cell is trigonal (space group
$P_{\bar{3}m1}$) with experimentally derived lattice parameters
$a=b=4.05$ \AA, and $c=6.74$ \AA. It contains 1 Mg, 2 Al, and 2 Si
atoms giving the formula \kj{MgAl_2Si_2}. The unit cell and atomic
positions are shown in table \ref{U1coordinates} and
fig.~\ref{U1new}. Experiments show that the \kj{<001>} direction
of the $U1$ unit cell runs parallel to the \kj{<310>} direction of
the fcc Al matrix \cite{U1experimental}. The $U1$ phase usually forms rod-like
precipitates with a length of 50-500 nm and a width of
approximately 50 nm. Performing a volume relaxation, we found the
calculated unit cell equilibrium volume to differ by only 1\% from
the experimentally derived unit cell volume. From a second order
Birch fit we obtained a bulk modulus of 71 GPa.

\begin{figure}
\rotatebox{00.00}{\scalebox{0.35}{\includegraphics{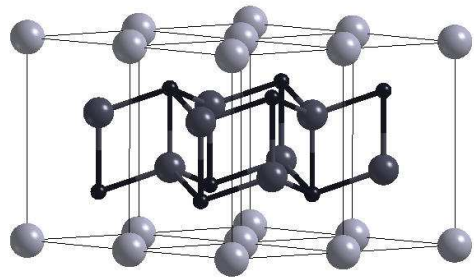}}}
\caption{\label{U1network} $U1$ bonding network}
\end{figure}

\begin{figure}
\rotatebox{00.00}{\scalebox{0.85}{\includegraphics{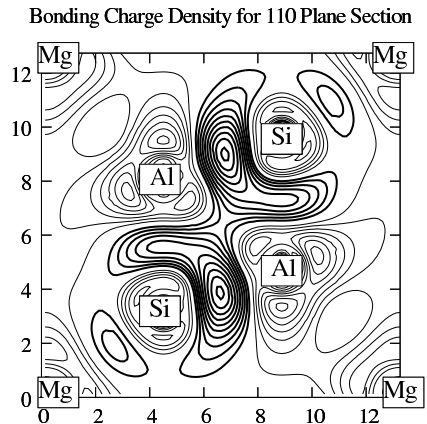}}}
\caption{\label{rhoU1}$U1$ bonding charge density for the (110)
plane. Thick lines represent contour levels with a positive value,
thin lines a negative value.}
\end{figure}

The $U1$ phase can be categorized as belonging to a class of
structures given the name \kj{CaAl_2Si_2}-type Zintl compounds
\cite{CaMg2Al2}. In Zintl compounds, the electropositive elements
are thought of as merely electron donors to the electronegative
elements which thereby are able to fulfill the octet rule. For
\kj{MgAl_2Si_2} this implies that each Mg atom donates two
electrons to \kj{Al_2Si_2} for each unit cell. This gives
\kj{Al_2Si_2^{2-}}, or two units of \kj{AlSi^-} which is
isoelectronic to AlN, fulfilling the octet rule. The resulting
layered structure is sketched in fig.~\ref{U1network}. \kj{AlSi^-}
constitutes a double layer of tightly bound ``chair-like''
six-membered rings separated by layers of hexagonal \kj{Mg^{2+}}.
In the Al-Si network, each Si atom bonds to 4 Al atoms forming an
umbrella-like structure, while each Al atom forms the more common
tetrahedral structure with 4 nearest neighbour Si atoms The Al-Si
bond length within each AlSi layer is 2.48 \AA, while the length
of the Al-Si bonds connecting the two \kj{AlSi^-} layers is 2.62
\AA. The length of the Mg-Si bond is 2.86 {\AA } (see below and
fig. \ref{rhoU1}), somewhat larger than the length of the partly
ionic bond in the $\beta$ equilibrium structure, but similar to
Mg-Si bonds found in the \kj{Mg_5Si_6} $\beta''$ phase
\cite{peter}.

The calculated bonding charge density for the (110) plane is shown
in fig~\ref{rhoU1}. There is a clear concentration of charge
between the Al and Si atoms indicating strong Al-Si bonding. There
is also a small build-up of charge between the Si and corner Mg
atoms which constitutes a coupling between Mg\kjo{2+} and
AlSi\kjo{-} layers.

\begin{figure}
\begin{tabular}{c}
\mbox{a)}{\rotatebox{-90.00}{\scalebox{0.30}{\includegraphics{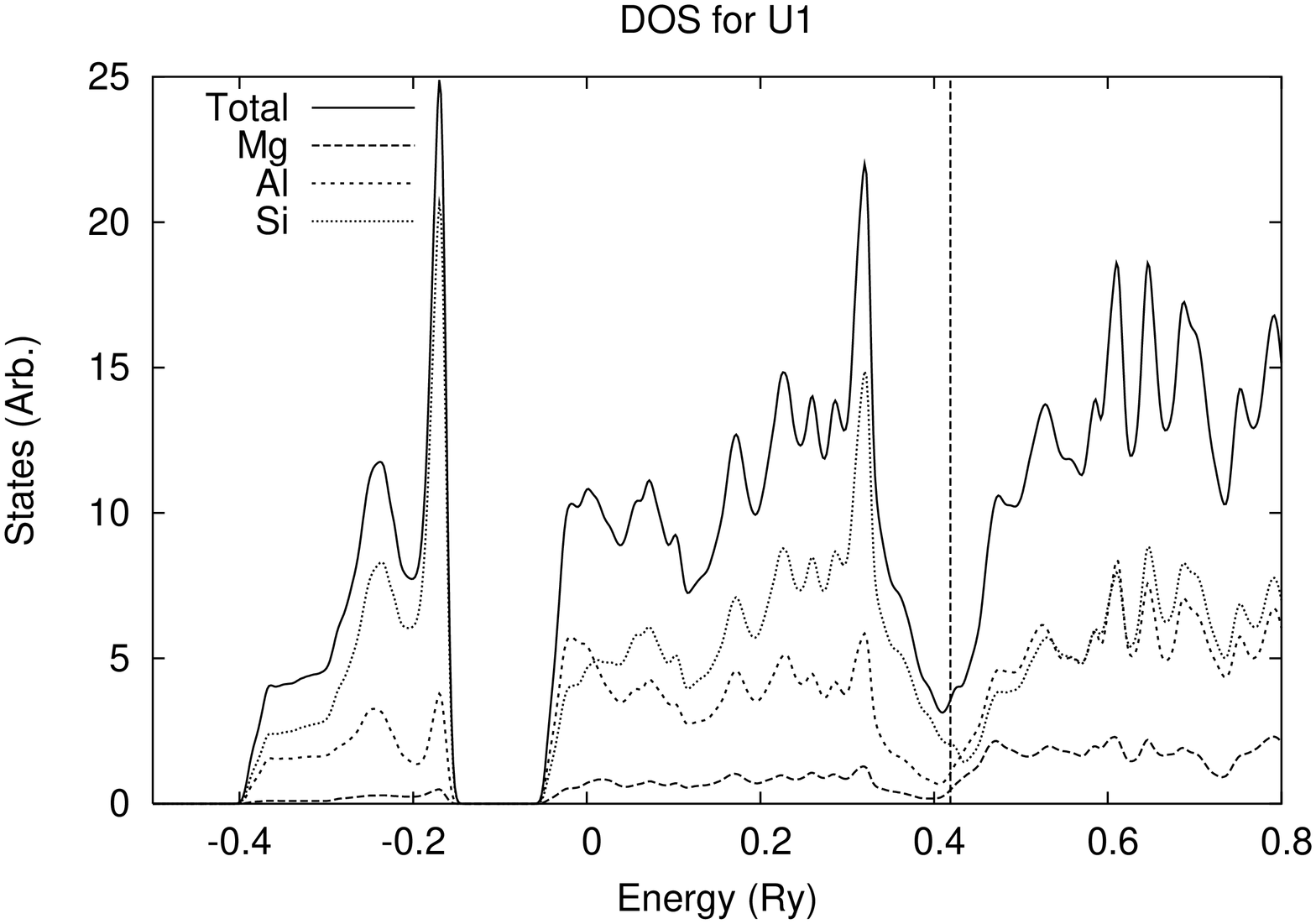}}} } \\
\mbox{b)}{\rotatebox{-90.00}{\scalebox{0.30}{\includegraphics{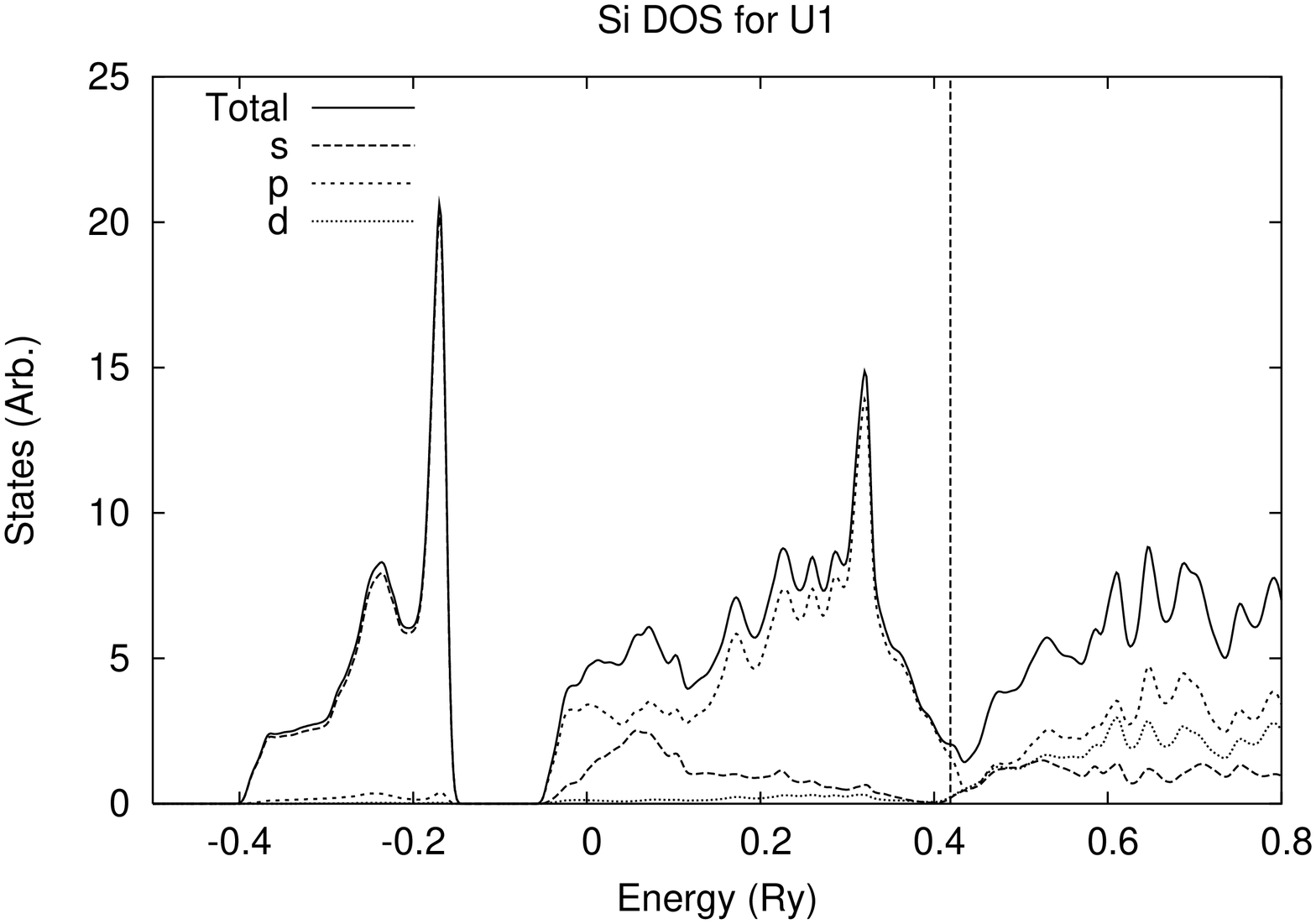}}} }
\end{tabular}
\caption{\label{DOSU1_1}a) total and b) Si muffin tin decomposed
DOS for the $U1$ phase. Note that the Al and Si DOS consists of
the contribution from 2 symmetrically equivalent atoms, while the
Mg DOS stems from only one atom. One therefore has to divide by a
factor two to get the atomic DOS for Al and Si.}
\end{figure}

\begin{figure}
\begin{tabular}{c}
\mbox{a)}{\rotatebox{-90.00}{\scalebox{0.30}{\includegraphics{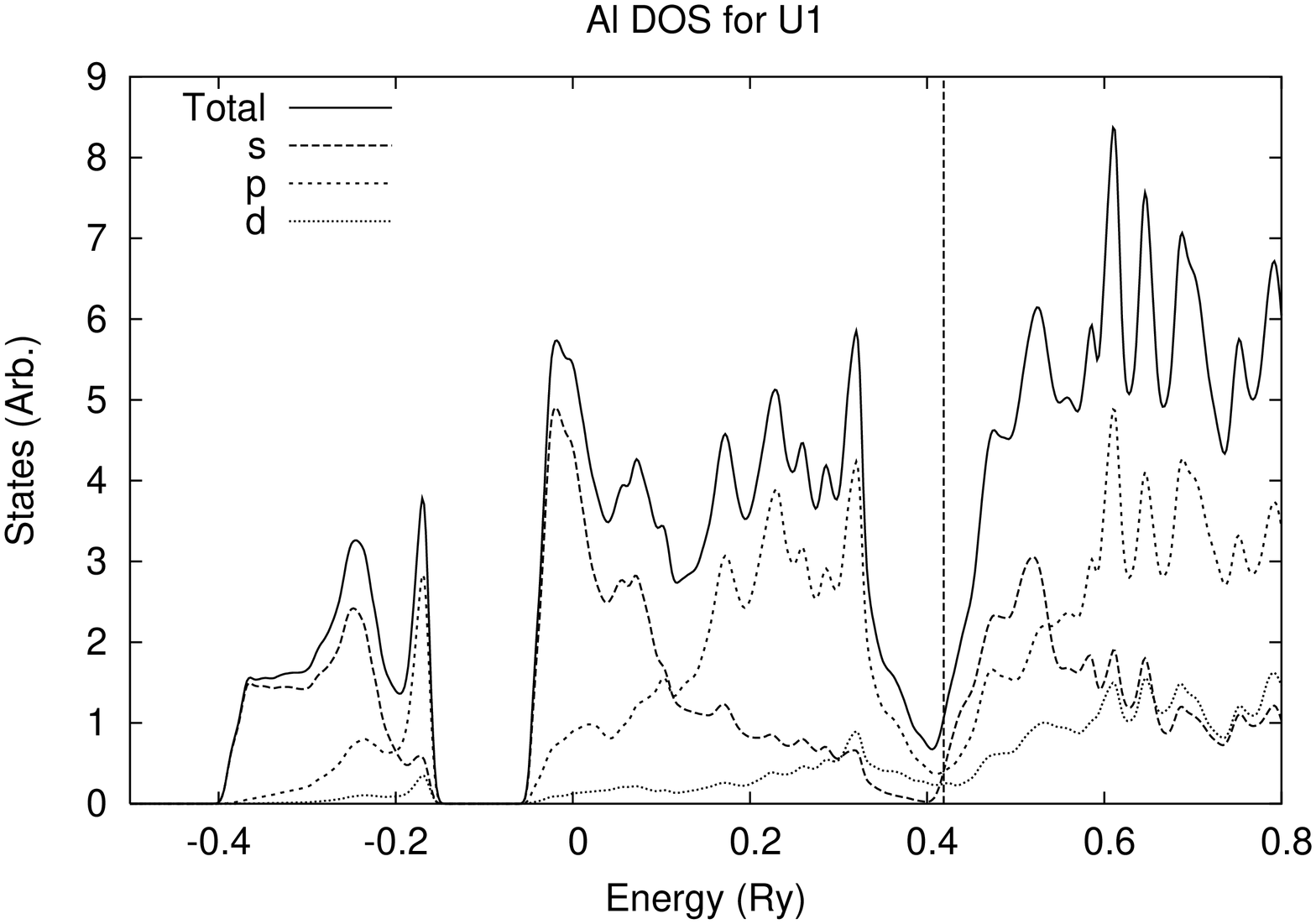}}} } \\
\mbox{b)}{\rotatebox{-90.00}{\scalebox{0.30}{\includegraphics{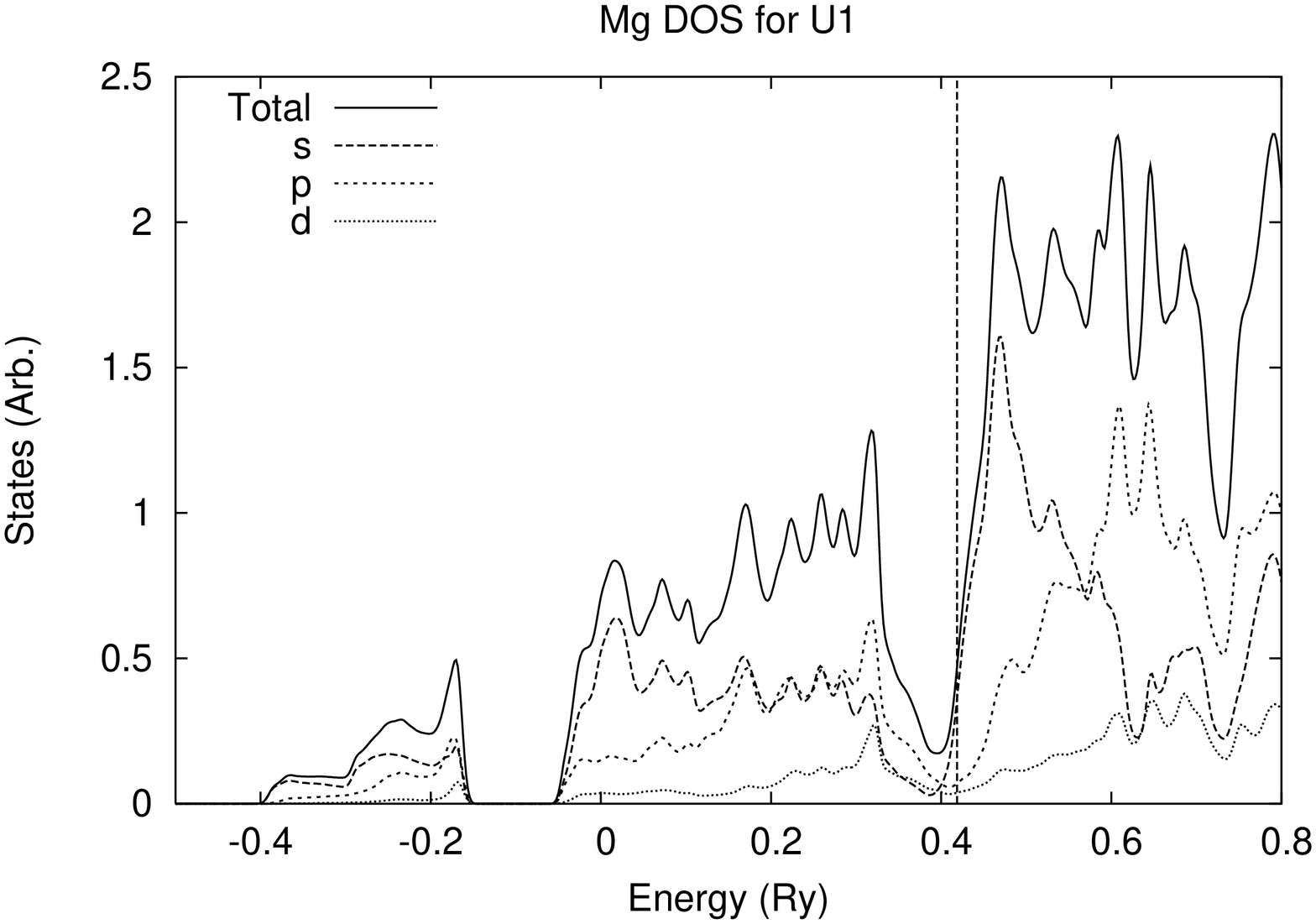}}} }
\end{tabular}
\caption{\label{DOSU1_2}a) Al and b) Mg muffin tin decomposed DOS for the U1 phase.}
\end{figure}

Evidence of the layered structure can also found in the DOS.
Figs.~\ref{DOSU1_1} and \ref{DOSU1_2} show the muffin tin
decomposed total and partial DOS. The total DOS below the fermi
level is characterized by two broad bands, separated by a 0.1 Ry
band gap (fig.~\ref{DOSU1_1}a). It is also evident that the major
contribution to the total DOS stems from the Al and Si states.
This is an indication that the Mg donates charge to the Al-Si
network and is not strongly involved in the electronic bonding.
Furthermore, the Si states give a relatively larger contribution
to the total DOS than the Al states. From the angular decomposed
Si DOS (fig.~\ref{DOSU1_1}b) one can see that the contribution to
the lower band comes mainly from Si $s$ states, while the higher
band consists mainly of $p$ states with a modest mixing in of
$s$-states. This can be seen as an indication of a partial
hybridization of the Si states resulting in the Si bonding
environment being partly covalent. This hybridization can also be
identified in the Al DOS (fig.~\ref{DOSU1_2}a). Compared to a
metallic DOS \cite{kittel}, there is a peak concentration of $s$
states with mixed in $p$ states slightly below 0 Ry resembling the
\kj{L_1} lobe of a covalent DOS. However for both Al and Mg
(fig.~\ref{DOSU1_2}b), the overall shape of the total atomic DOS
seems roughly to follow the square-root relationship
characteristic of a metallic DOS.

The origin of the band gap at the centre of the occupied states
can be understood from the same line of reasoning as for the
$\beta$ phase. In the $U1$ phase, the Si atoms are arranged as an
hcp lattice with lattice parameters $a=4.05$ {\AA} and $c=6.74$
{\AA}. If we once again consider only the Si system (with a
nearest neighbour distance of 4.05 {\AA}), then little
hybridization of the atomic Si $s$ and $p$ valence states can be
expected and a similar DOS to the $\beta$ is seen. That is, the
occupied valence states will be separated into two relatively
narrow bands. By introducing the Al and Mg, the Al forms a tightly
bound bonding network with the silicon via the donation of charge
from the Mg atoms. This coupling results in a broadening (with
respect to the artificial Si sub-structure) of the two occupied
bands and in the removal of the band gap seen in $\beta$. This has
its origins in the stronger mixed $s$ and $p$ character of the
second occupied band (centred at $\approx-0.2$ Ry) of Si
(fig.~\ref{DOSU1_1}b), when compared to the strong $p$ character
of the corresponding band for $\beta$ (fig.~\ref{Mg2Si_DOS2}) ---
indicating strong hybridization between the Si $p$ states and the
Al $s$ and $p$ states.

%

\subsection{$U2$}

\begin{figure}
\begin{tabular}{c}
\mbox{a)}{\rotatebox{00.00}{\scalebox{0.35}{\includegraphics{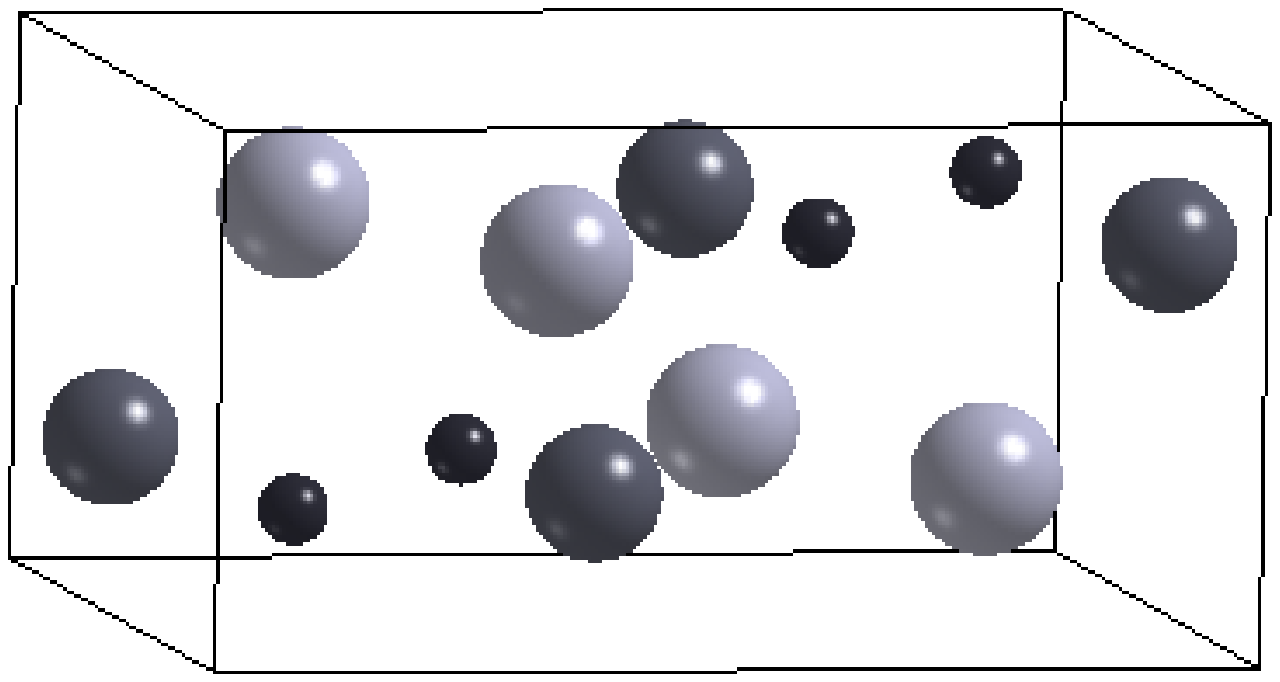}}} } \\
\mbox{b)}{\rotatebox{00.00}{\scalebox{0.35}{\includegraphics{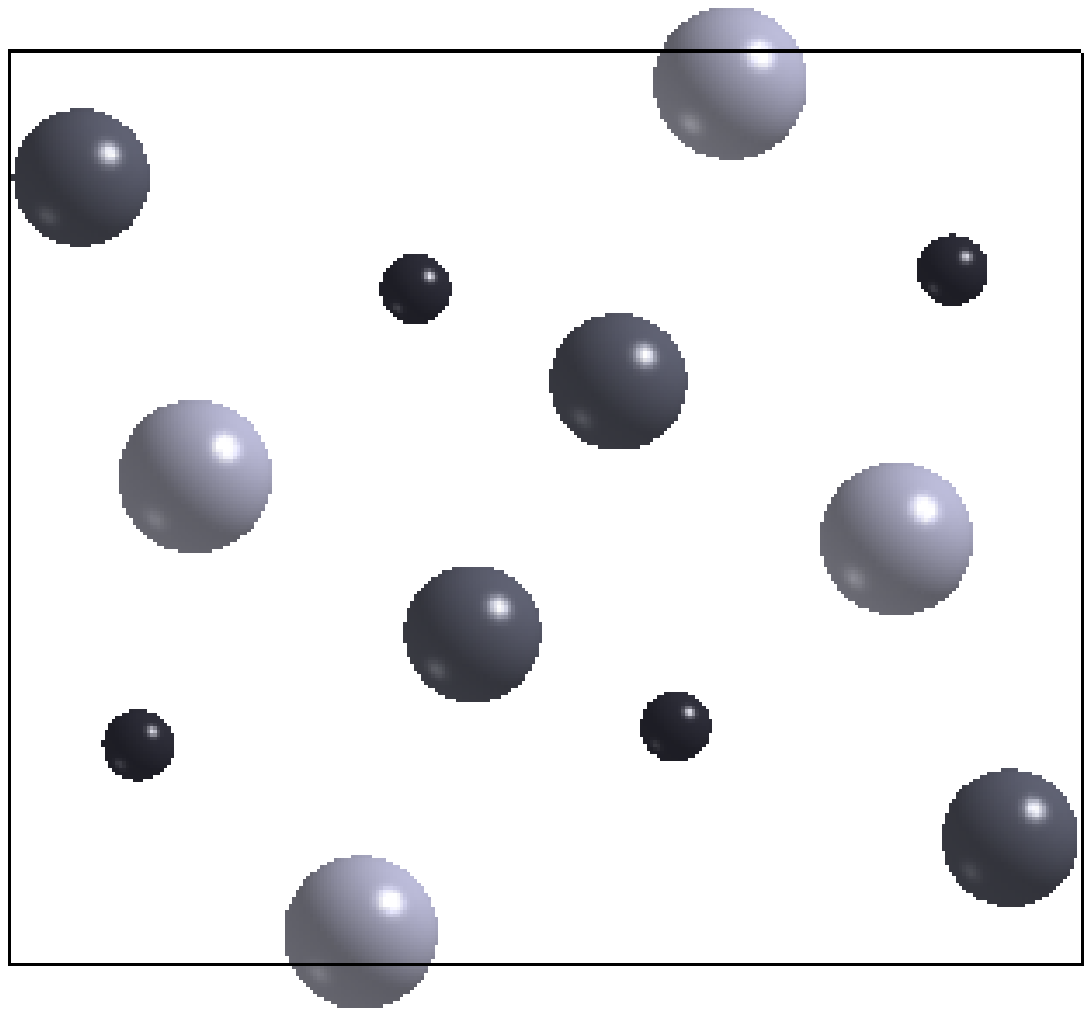}}} }
\end{tabular}
\caption{\label{U2new} U2 conventional unit cell in a) perspective and b) \kj{<010>} direction}
\end{figure}

The $U2$ primitive/conventional unit cell is orthorhombic (space
group $P_{nma}$) with experimentally derived lattice parameters
$a=6.75\AA$, $b=4.05\AA$, and $c=7.94\AA$. It contains 4 Mg, 4 Al,
and 4 Si atoms giving the formula \kj{Mg_4Al_4Si_4} (fig.~\ref{U2new}). The
coordinates of the symmetrically inequivalent atoms of the unit
cell are shown in table \ref{U2coordinates}. The morphology and
size of the precipitate formed by the $U2$ phase is usually the
same as for the $U1$ phase. The \kj{<100>} direction of the unit
cell is oriented parallel to the \kj{<310>} direction of the fcc
Al matrix\cite{U2experimental}. By performing a volume relaxation of the unit cell, we
get an equilibrium volume differing by less than 1\% from the
experimentally derived unit cell volume. A second order Birch fit
gives a bulk modulus of 69.1 GPa, very close to the value obtained
for the $U1$ phase.

\begin{figure}
\rotatebox{00.00}{\scalebox{0.35}{\includegraphics{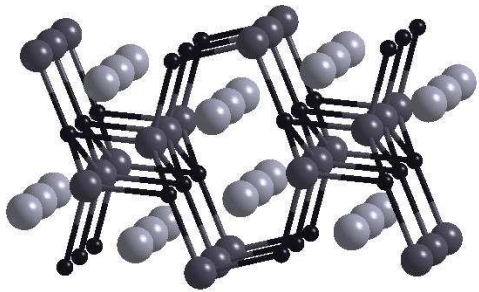}}}
\caption{\label{U2network} U2 bonding network}
\end{figure}

The $U2$ phase is similar to the TiNiSi structure type
\cite{TiNiSi, TiNiSibonding}. As for the $U1$ phase, the bonding
in the $U2$ phase is characterized by the electropositive Mg atoms
donating charge to the electronegative Al and Si atoms forming a
tightly bound bonding network (fig.~\ref{U2network}). However,
unlike for the $U1$ phase, the electropostitive Mg atoms retain a
large amount of charge and (using a simple chemical picture) there
is not enough transfer for the electronegative elements to fulfill
the octet rule \cite{TiNiSibonding}. The Al-Si bond lengths are
all $2.59$ to $2.61\AA$, while the bond lengths of the Mg-Si
neighbour coupling is $2.78$ and $2.86\AA$.
\begin{table}
\caption{\label{U2coordinates} Fractional coordinates for the inequivalent atomic positions
of the U2 phase unit cell \cite{U2experimental}.}
\begin{ruledtabular}
\begin{tabular}{lccr}
\bf{Atom} & \bf{a} & \bf{b} & \bf{c} \\
\hline
Mg & 0.034(9) & 3/4 & 0.327(4) \\
Al & 0.361(4) & 1/4 & 0.432(5) \\
Si & 0.239(3) & 1/4 & 0.120(9) \\
\end{tabular}
\end{ruledtabular}
\end{table}

\begin{figure}
\begin{tabular}{c}
\mbox{a)}{\rotatebox{00.00}{\scalebox{0.75}{\includegraphics{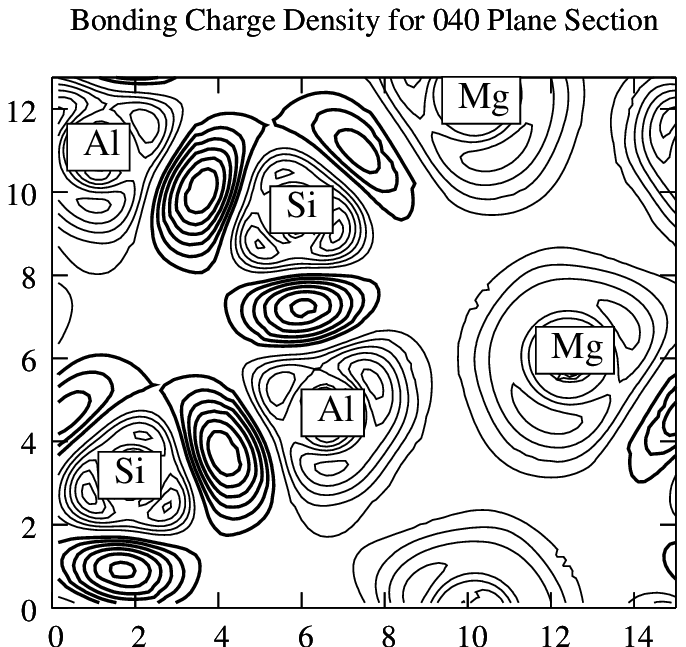}}} } \\
\mbox{b)}{\rotatebox{00.00}{\scalebox{0.75}{\includegraphics{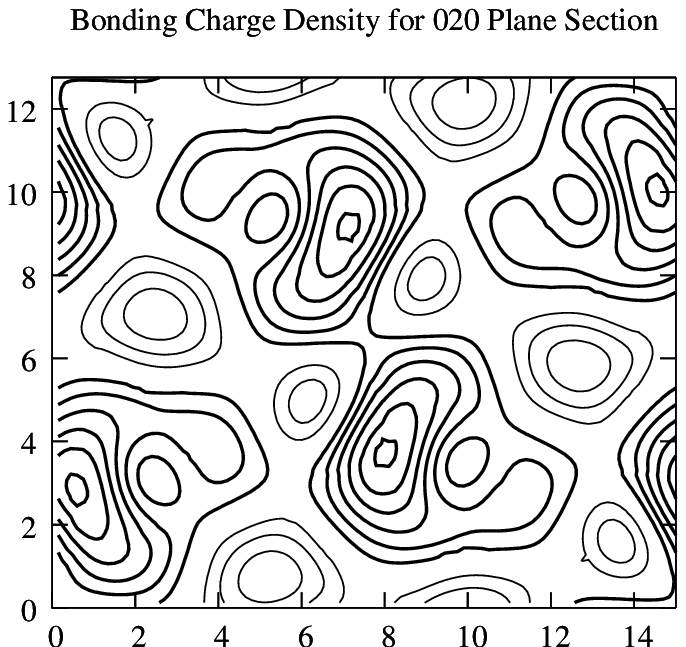}}} } \\
\mbox{c)}{\rotatebox{00.00}{\scalebox{0.75}{\includegraphics{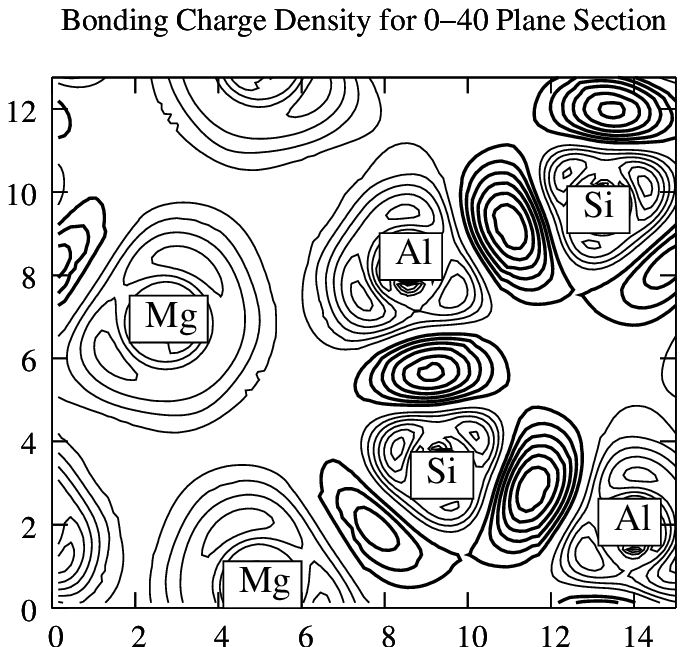}}} }
\end{tabular}
\caption{\label{rhoU2new}Bonding charge density for a) (040) plane
b) (020) plane and c) \kj{(0\bar40)} plane of the $U2$ unit cell. Thick lines
represent contour levels with a positive value, thin lines a
negative value.}
\end{figure}

Fig.~\ref{rhoU2new} displays the bonding charge density maps for
the (040), (020) and (0\kj{\bar4}0) planes of the $U2$ unit cell.
As for the $U1$ phase, there is a concentration of charge between
Al and Si nearest neighbours making up the AlSi bonding network.
This charge concentration appears to have more of a covalent,
directional character, than the $U1$ Al-Si bond. Furthermore, one
can identify a more pronounced charge concentration in between
Mg-Si neighbours than for the $U1$ phase. This fits well with the
general assumption for this structure class --- that a large
amount of Mg charge is retained. As for the U1 phase there is no
indication of charge build-up between Al and Mg neigbours,
indicating that the AlSi network is mainly coupled to the Mg atoms
via the Si atoms.
\begin{figure}
\begin{tabular}{c}
\mbox{a)}{\rotatebox{-90.00}{\scalebox{0.30}{\includegraphics{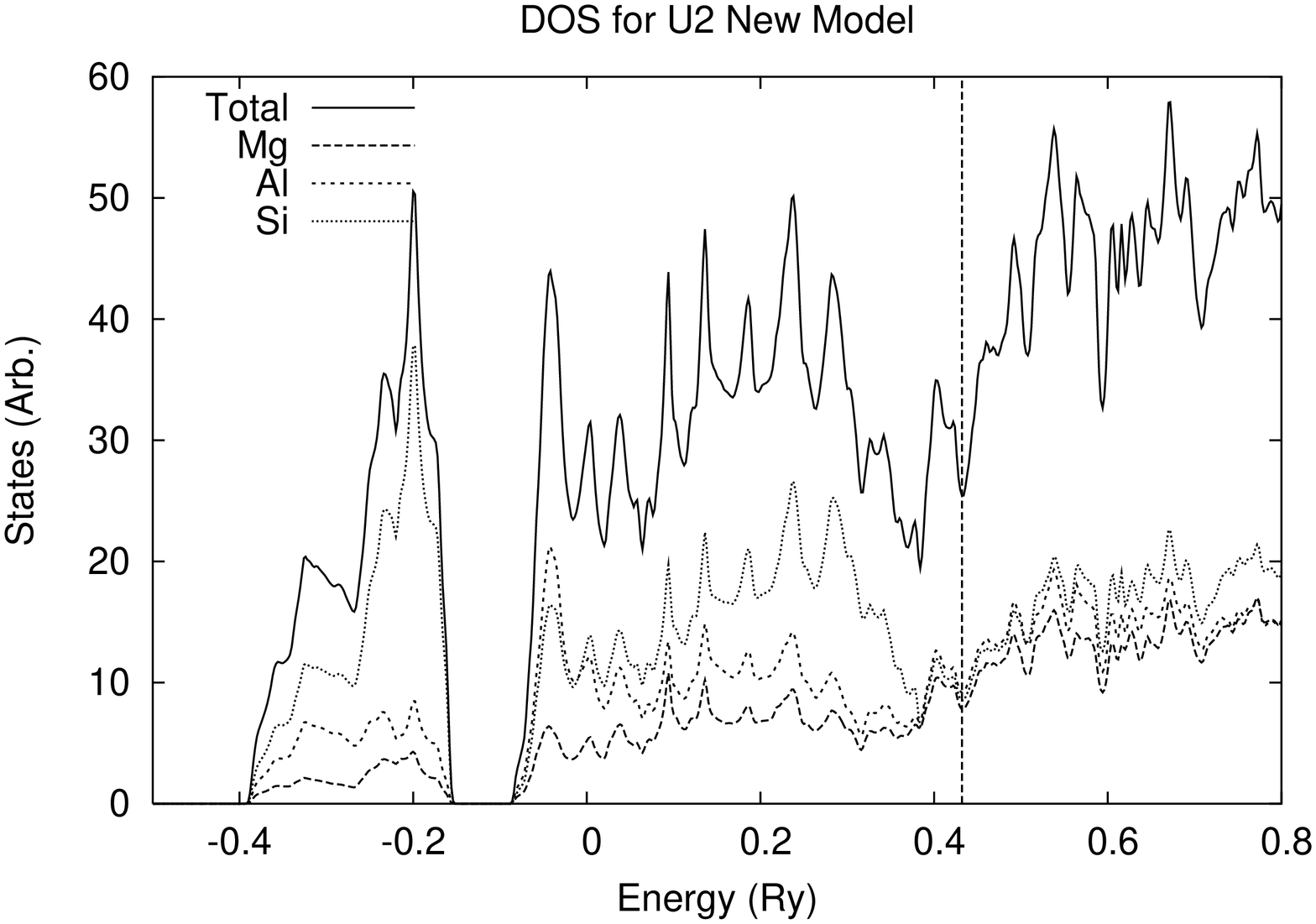}}} } \\
\mbox{b)}{\rotatebox{-90.00}{\scalebox{0.30}{\includegraphics{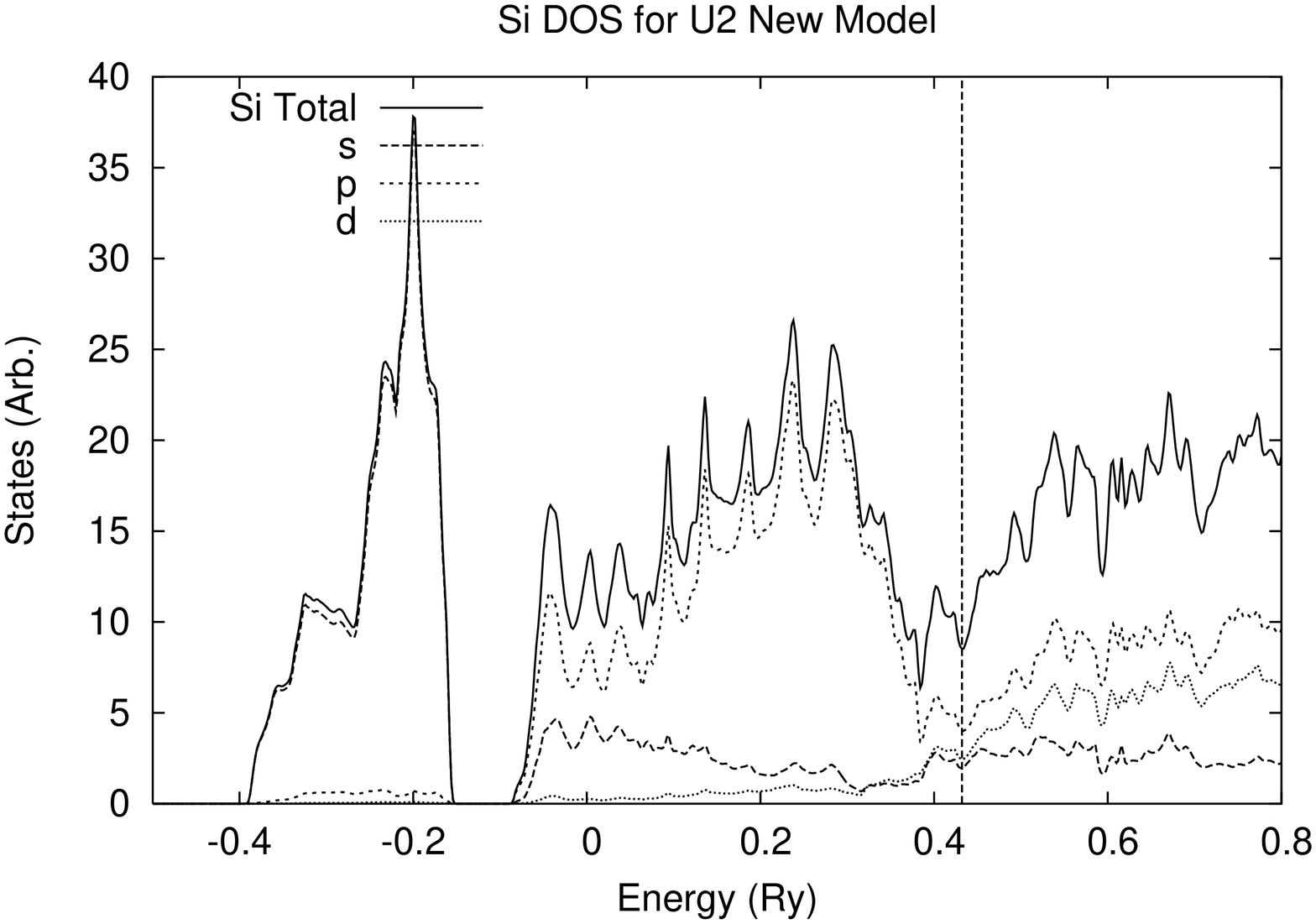}}} }
\end{tabular}
\caption{\label{DOSU2_1} a) total and b) Si muffin tin decomposed
DOS for the $U2$ phase.}
\end{figure}
\begin{figure}
\begin{tabular}{c}
\mbox{a)}{\rotatebox{-90.00}{\scalebox{0.30}{\includegraphics{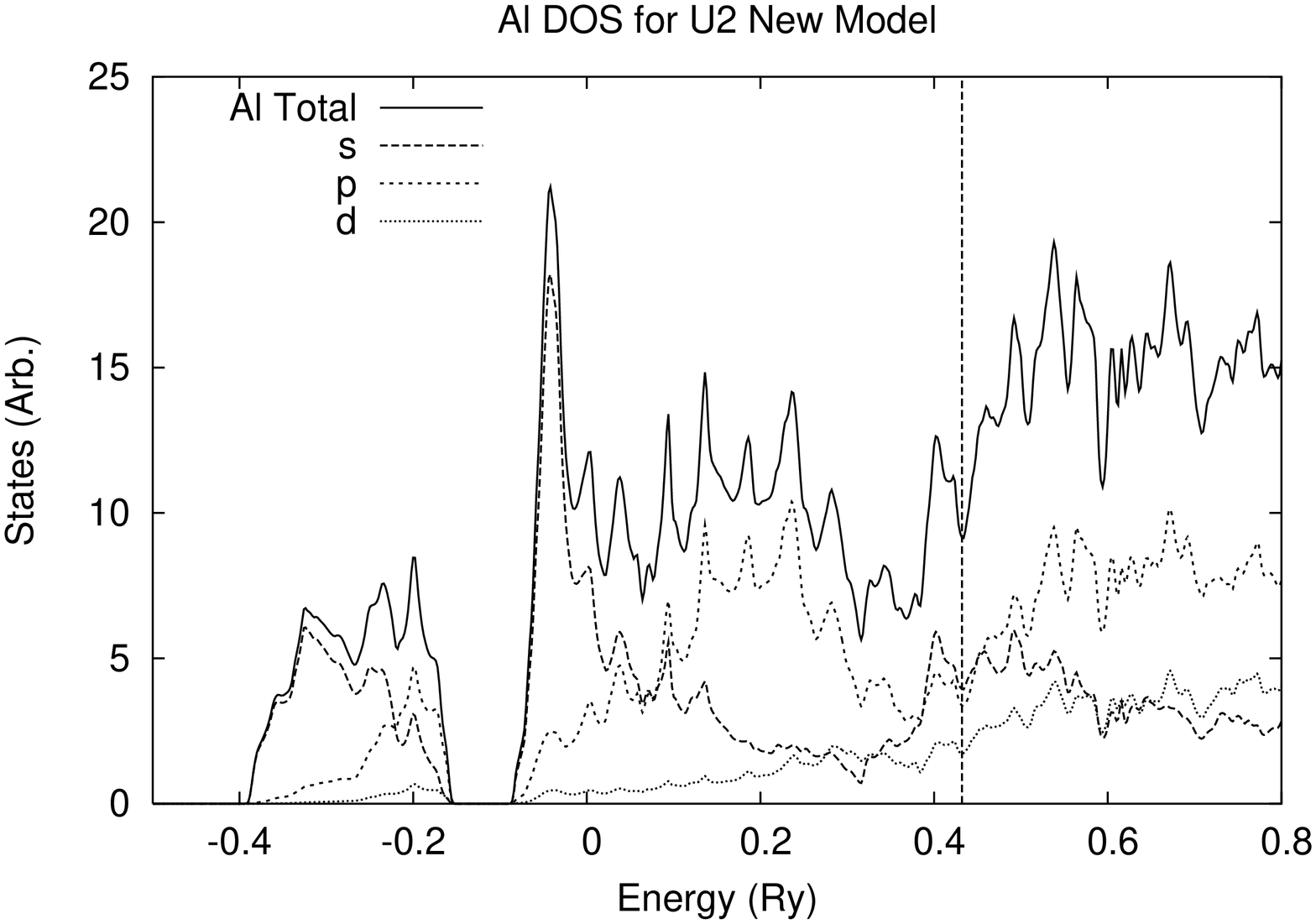}}} } \\
\mbox{b)}{\rotatebox{-90.00}{\scalebox{0.30}{\includegraphics{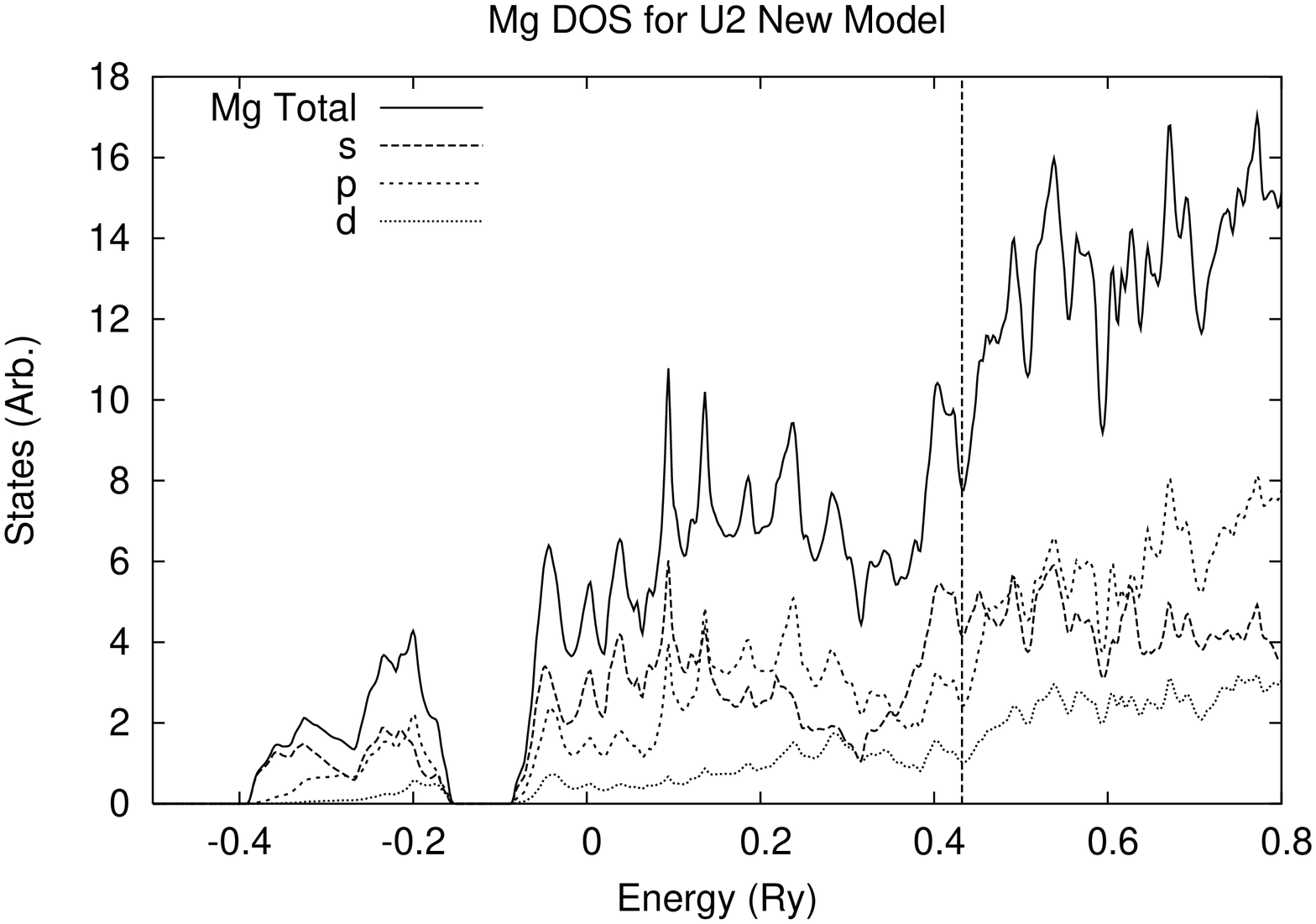}}} }
\end{tabular}
\caption{\label{DOSU2_2} a) Al and b) Mg muffin tin decomposed DOS
for the $U2$ phase.}
\end{figure}

This picture is further justified by the total and partial DOS
shown in figs.~\ref{DOSU2_1} and \ref{DOSU2_2}. Comparing the DOS
of the $U2$ phase with the DOS of the $U1$ phase one can identify
several common features, supporting the proposition that the
electronic bonding picture is very similar for the two phases.
Note that the total DOS is the sum of 1 Mg, 1 Al, and 1 Si atom,
as opposed to 1 Mg, 2 Al, and 2 Si atom for the $U1$ phase
(fig.~\ref{DOSU1_1}a). From the atom decomposed DOS of the $U2$
phase (fig. \ref{DOSU2_1}a) one can see that the Al and Si states
are dominating with respect to the Mg states, and that the Si
states give a slightly larger contribution than the Al states.
This is consistent with the donation of charge from Mg to Al+Si
atoms, as in the case of $U1$. However, the Mg DOS gives a larger
contribution to the total DOS than for the $U1$ phase. This is an
indication that there is less Mg to Al and Si charge transfer,
than for the $U1$ Zintl-type phase. We remind the reader that it
is difficult to quantify the charge transfer because the
(L)APW+(lo) method does not use an atomic/localised basis set.
Furthermore, the magnitude of the band gap separating the two
occupied bands is slightly smaller for the $U2$ phase than for the
$U1$ phase (0.075 Ry vs. 0.1 Ry for the $U1$ phase). Consequently,
the region of hybridization around 0 Ry is increased for the Al
and Si DOS (fig.~\ref{DOSU2_2} a and b) consistent with the
pronounced covalent character of the Al-Si bonds found in the
bonding charge density.

As for the $U1$ and $\beta$ phase, the origin of the band gap in
the occupied states can be explained by the separation of the Si
atoms with respect to each other, which for this structure forms a
slightly distorted h.c.p structure (not shown here) with lattice
parameters comparable to those of the $U1$ Si hcp sub-lattice.

\section{Discussion and Concluding remarks}

\begin{figure}
\centerline{\rotatebox{-90.00}{\scalebox{0.35}{\includegraphics{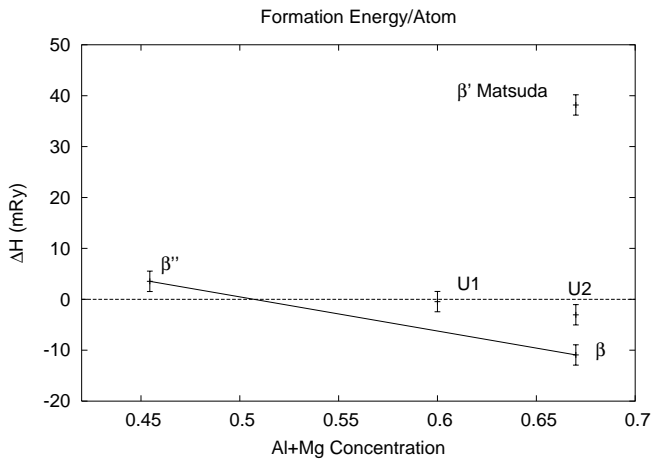}}}}
\caption{\label{AllCoh} Calculated energies of formation for the
various precipitate phases. The error bars indicate the estimated
level of accuracy of $\pm 2.0$ mRy. The line connecting $\beta''$
and $\beta$ is a guide to the eye.}
\end{figure}

We have made a comparative study of the electronic structure of
the $\beta$, $\beta''$, $U1$, and $U2$ phases in the AlMgSi alloy
precipitation sequence. The bonding in the $\beta$ phase is
characterized by the covalent bonding between Si-Si nearest
neighbour pairs and ionic/covalent bonding between Si-Mg nearest
neighbour pairs. For the $\beta$ phase the bonding is dominated by
the partly ionic Mg-Si bond. By calculating the heat of formation
for the various precipitate phases further insight can be gained
in the observed precipitation sequence. Presently we employ the
formula:
\begin{equation}
\Delta H = E_{\text{AlMgSi}} - x_{\text{Mg}} E_{\text{Mg}} - x_{\text{Si}} E_{\text{Si}} - x_{\text{Al}} E_{\text{Al}},
\end{equation}
where $\Delta H$ is the formation energy/enthalpy per atom,
$E_{\text{AlMgSi}}$ is the energy of the given AlMgSi compound,
and  $E_{\text{Mg}}$, $E_{\text{Si}}$, and $E_{\text{Al}}$ are the
equilibrium ground state (zero temperature) energies per atom at
of hcp Mg, dc Si, and fcc Al respectively. $x_{i}$ is the relative
content of element $i$ in the compound. Results for the various
phases in the precipitation sequence are shown in table
\ref{Formation Energies} and figure \ref{AllCoh}. For the $\beta$
and $\beta''$ phases we used the experimentally observed lattice
parameters. For the $\beta''$ we used the relaxed coordinates
found from force minimization (table \ref{betadpcoordinates}). We
have also included the corresponding energy from a past model for
the $\beta'$ phase proposed by Matsuda et al.\cite{matsuda}.

\begin{table*}
\caption{\label{Formation Energies} Calculated formation energies
and bulk moduli. Space groups are given with space group number in
parentheses.}
\begin{ruledtabular}
\begin{tabular}{lccccr}
Structure & Space & Al:Mg:Si & Lattice & Bulk Modulus & Energy/Atom \\
Type & Group & in Unit Cell & Parameters (\AA) & (GPa) & (mRy) \\
\hline
$\beta''$\footnotemark[1] & C2/m (12) & 0:5:6 & a=15.16 b=6.74 c=4.05 & 65  & 3.53 \\
$\beta$  & Fm$\bar{3}$m (225) & 0:4:2 & a=b=c=6.39 & 54 & -10.93 \\
$\beta'$ Matsuda\footnotemark[2] & P$\bar{6}$2m (189) & 0:4:2 & a=b=7.1 c=4.05 & NA  & 38.174 \\
$U2$ & $P_{\bar{3}m1}$ (164) & 4:4:4 & a=b=4.05 c=6.74 & 69  & -3.04 \\
$U1$  &  $P_{nma}$ (62) & 2:1:2 & a=6.75 b=4.05 c=7.94 & 71  & -0.45 \\
\end{tabular}
\end{ruledtabular}
\footnotetext[1]{Ref. \onlinecite{science}} \footnotetext[2]{Ref.
\onlinecite{matsuda}}
\end{table*}

It is difficult to draw any quantitative conclusions based on the
relative formation energy of these phases in Aluminium, at finite
temperatures, without taking into account the effects of the
interface and entropy. The calculated energies are nonetheless
very reasonable. As can be seen, only 3 phases ($\beta$, $U1$ and
$U2$), give a negative formation energy. The $U1$ and $U2$ are
lower in energy than $\beta''$, but higher than the $\beta$ phase.
$\beta$ shows the most negative energy, which is line with
expectations, since it is the terminal equilibrium structure of
the precipitation sequence. The energy of $\beta''$ is slightly
positive, but not markedly different from $\beta$, $U1$ and $U2$.
$\beta''$, which can be derived from an fcc super-cell, forms 
early in the precipitation sequence at
temperatures above 150 $^\circ$C, but gradually disappears after
further heat treatment \cite{calin}.

We note the unusually high heat of formation for Matsuda's
$\beta'$ model which has a hexagonal unit cell with space group
\kj{P\bar62m} with experimentally derived lattice parameters
$a=b=4.05{\AA}$, $c=6.74{\AA}$ and the stoichiometry \kj{Mg_2Si}.
For the calculated formation energy, we used the experimentally
derived lattice parameters and published coordinates. Upon volume
relaxation, we however found that the structure reduced in volume
by 26\% corresponding to a heat of formation of 25 mRy. Furthermore,
from a second order Birch fit, we find that for the experimental
lattice constant, the Matsuda's $\beta'$ phase is under a negative
pressure of approximately 6.9 GPa. Since the experimentally
derived unit cell parameters are regarded as being extremely
accurate it becomes difficult to see how such as phase might exist
within the Al matrix.

To this date $U1$ and $U2$ are the only phases in the
precipitation sequence of the AlMgSi alloy system containing
Aluminium in addition to Magnesium and Silicon. The
crystallographic structure type of both phases belong to structure
families containing an abundance of documented compounds. A
characteristic for both phases is the formation of a tightly bound
network of Al and Si atoms made possible by at charge transfer
from the electropositive Mg atoms. Considering the wide range of
structures belonging to this family it is possible that these
these type of structures containing other hardening elements than
Mg and Si would form stable precipitates in Aluminium alloys. For
example, it is quite remarkable that the $U1$ phase with Magnesium
replaced by the rare earth element Europium produces a
\kj{CaAl_2Si_2}-type structure with lattice parameters and
coordinates differing from the $U1$-phase by only few percent
\cite{lesscommon}. Suggested further work include the calculations
on these two phases with iso-valent species substituted for the Si
and Mg sites. In the future these calculations could also be used
together with interfacial energies and entropy calculations to
estimate the solid solubility of these phases in Aluminium.

Furthermore, the rule of thumb that successful aluminium
precipitation hardened alloys have secondary and ternary elements
that are larger and smaller than aluminium is not sufficiently
accurate when applied to the stability of bulk phases in the
precipitation sequence. The atomic radii of Al, Si, and Mg vary
with the bonding environment. For example, in the equilibrium
$\beta$ phase the atomic size of Si is increased and that of Mg
correspondingly decreased in forming a dominating ionic/covalent
Mg-Si bond. Also, using the appropriate values for ionic radii
\cite{kittel}, the Mg-Si bond length in the $\beta$ and $\beta''$
phases are predicted to be considerably smaller than what the
present calculations show. A more precise formulation would
involve the ability of the secondary(A) and ternary(B) elements to
form stable bonds (A-B, A-A, and B-B) with bond lengths which are
comparable that of the Al-Al bond.

\begin{acknowledgments}
This study has been performed as part of the research programs 
``KMB Heat Treatment Fundamentals'' and ``SUP Micro- and Nanostructure 
Based Materials Development'', supported by the Norwegian Research 
Council and collaborators from the industry. Parts of this work has 
been supported by the Norwegian Research Council through CPU time
on the NOTUR Origin 3800.
\end{acknowledgments}

\bibliography{AlMgSi}

\end{document}